\documentclass[aps,pre,twocolumn,superscriptaddress,floatfix]{revtex4-1}

\usepackage{graphicx,bm,amssymb,mathrsfs,amsmath,dcolumn,hyperref}
\usepackage{multirow}
\usepackage{enumerate}
\usepackage{enumitem}
\usepackage{color}
\usepackage{subfigure}  
\usepackage{epstopdf}
\usepackage{bbm}
\usepackage{threeparttable}
\usepackage{mathtools}
\usepackage{tikz}
\usepackage{tikz-3dplot} 
\usepackage{float}
\usepackage{indentfirst}
\usetikzlibrary{math}
\usepackage{microtype}

\usepackage[utf8]{inputenc} 
\usepackage[greek, main=english]{babel}

\begin{document}
\newcommand{\upcite}[1]{\textsuperscript{\textsuperscript{\cite{#1}}}}
\newcommand{\be}{\begin{equation}}
\newcommand{\ee}{\end{equation}}
\newcommand{\half}{\frac{1}{2}}
\newcommand{\ith}{^{(i)}}
\newcommand{\im}{^{(i-1)}}
\newcommand{\gae}
{\,\hbox{\lower0.5ex\hbox{$\sim$}\llap{\raise0.5ex\hbox{$>$}}}\,}
\newcommand{\lae}
{\,\hbox{\lower0.5ex\hbox{$\sim$}\llap{\raise0.5ex\hbox{$<$}}}\,}

\definecolor{blue}{rgb}{0,0,1}
\definecolor{red}{rgb}{1,0,0}
\definecolor{green}{rgb}{0,1,0}
\newcommand{\blue}[1]{\textcolor{blue}{#1}}
\newcommand{\red}[1]{\textcolor{red}{#1}}
\newcommand{\green}[1]{\textcolor{green}{#1}}
\newcommand{\orange}[1]{\textcolor{orange}{#1}}
\newcommand{\yd}[1]{\textcolor{blue}{#1}}

\newcommand{\scrA}{{\mathcal A}}
\newcommand{\scrE}{{\mathcal E}} 
\newcommand{\scrF}{{\mathcal F}} 
\newcommand{\scrL}{{\mathcal L}}
\newcommand{\scrM}{{\mathcal M}} 
\newcommand{\scrN}{{\mathcal N}}
\newcommand{\scrS}{{\mathcal S}}
\newcommand{\scrs}{{\mathcal s}}
\newcommand{\scrP}{{\mathcal P}}
\newcommand{\scrO}{{\mathcal O}}
\newcommand{\scrR}{{\mathcal R}}
\newcommand{\scrC}{{\mathcal C}}
\newcommand{\scrV}{{\mathcal V}}
\newcommand{\scrD}{{\mathcal D}}
\newcommand{\scrG}{{\mathcal G}}
\newcommand{\scrW}{{\mathcal W}}
\newcommand{\PP}{\mathbb{P}}
\newcommand{\ZZ}{\mathbb{Z}}
\newcommand{\EE}{\mathbb{E}}
\renewcommand{\d}{\mathrm{d}}
\newcommand{\dm}{d_{\rm min}}
\newcommand{\rhojunction}{\rho_{\rm j}}
\newcommand{\rhojunctionLim}{\rho_{{\rm j},0}}
\newcommand{\rhobranch}{\rho_{\rm b}}
\newcommand{\rhobranchLim}{\rho_{{\rm b},0}}
\newcommand{\rhononbridge}{\rho_{\rm n}}
\newcommand{\rhononbridgeLim}{\rho_{{\rm n},0}}
\newcommand{\percolationCluster}{C}
\newcommand{\leafFreeCluster}{C_{\rm \ell f}}
\newcommand{\bridgeFreeCluster}{C_{\rm bf}}
\newcommand{\df}{d_f}
\newcommand{\yt}{y_{\rm t}}
\newcommand{\yh}{y_{\rm h}}
\newcommand{\yu}{y_{\rm u}}
\newcommand{\dfprime}{d'_{\rm f}}
\newcommand{\yhhat}{\hat{y}_{\rm h}}
\newcommand{\ythat}{\hat{y}_{\rm t}}
\newcommand{\yhstar}{y^*_{\rm h}}
\newcommand{\ytstar}{y^*_{\rm t}}
\newcommand{\zc}{z_{\rm c}}
\newcommand{\dc}{d_{\rm c}}
\newcommand{\bfx}{{\bf x}}
\newcommand{\bfO}{{\bf o}}
\newcommand{\bfo}{{\bf o}}
\newcommand{\bfS}{{\bf S}}
\newcommand{\bfq}{{\bf q}}
\newcommand{\bfr}{\bf r}
\newcommand{\origin}{\bf 0}
\newcommand{\bfe}{\bf e}
\newcommand{\bfk}{{\bf k}}
\newcommand{\bfy}{\bf y}
\newcommand{\bfu}{\bf u}
\newcommand{\bmomega}{{\bm \omega}}
\newcommand{\bfU}{{\bf u}}
\newcommand{\ind}{\mathbbm{1}}
\newcommand{\xiu}{\xi_{\rm u}}

\newcommand{\Lm}{L_{\rm min}}
\newcommand{\scrT}{{\mathcal T}}

\title{High-precision Dynamic Monte Carlo Study of Rigidity Percolation}
\date{\today}
\author{Mingzhong Lu}
\thanks{These authors contributed equally to this work.}
\affiliation{Department of Modern Physics, University of Science and Technology of China, Hefei, Anhui 230026, China}
\author{Yufeng Song}
\thanks{These authors contributed equally to this work.}
\affiliation{Department of Modern Physics, University of Science and Technology of China, Hefei, Anhui 230026, China}
\author{Qiyuan Shi}
\affiliation{Department of Modern Physics, University of Science and Technology of China, Hefei, Anhui 230026, China}
\author{Ming Li}
\email{lim@hfut.edu.cn}
\affiliation{School of Physics, Hefei University of Technology, Hefei, Anhui 230009, China}
\author{Youjin Deng}
\email{yjdeng@ustc.edu.cn}
\affiliation{Department of Modern Physics, University of Science and Technology of China, Hefei, Anhui 230026, China}	
\affiliation{Hefei National Research Center for Physical Sciences at the Microscales, University of Science and Technology of China, Hefei 230026, China}	
\affiliation{Hefei National Laboratory, University of Science and Technology of China, Hefei 230088, China}	

\begin{abstract} 
Rigidity percolation provides an important basis for understanding the onset of mechanical stability in disordered materials. While most studies on the triangular lattice have focused on static properties at fixed bond~(site) occupation probabilities, the dynamics of the rigidity transition remain less explored. In this work, we formulate a dynamic pebble game algorithm that monitors how rigid clusters emerge and evolve as bonds are added sequentially to an empty lattice, with computational efficiency comparable to the standard static pebble game. We uncover a previously overlooked temporal self-similarity exhibited in multiple quantities, including the cluster size changes and merged cluster sizes during bond addition, as well as the number of simultaneously merging clusters. We identify large-scale cascade events in which a single bond addition triggers the merger of an extensive number of clusters that scales with system size with inverse correlation-length exponent. Using an event-based ensemble approach, we obtain high-precision estimates of the critical point $p_c = 0.660\,277\,8(10)$, the inverse correlation-length exponent $1/\nu = 0.850(3)$, and the fractal dimension $d_f = 1.850(2)$, representing substantial improvements over existing values.

\end{abstract}
\maketitle

\section{Introduction}
\label{sec:Introduction}
Rigidity percolation describes the emergence of mechanical stability in disordered networks through the formation of system-spanning rigid clusters. The transition occurs when constraints are sufficient to eliminate all zero-frequency vibrational modes, transforming a mechanically floppy network into one capable of supporting stress. This phenomenon governs critical behavior across diverse physical systems, from network glasses~\cite{Thorpe_2000} and colloidal gels~\cite{Broedersz_2011,zhang2019correlated} to biological tissues~\cite{petridou2021rigidity, rozman2024basolateral} and amorphous solids~\cite{zaccone2011approximate}. Despite extensive applications, a complete understanding of the critical behavior remains elusive, primarily due to anomalously strong finite-size scaling corrections that persist to very large system sizes.

In two-dimensional (2D) systems, Bernoulli percolation (BP), where bonds or sites are occupied independently with probability $p$, is well understood through exact solutions, with exact critical exponents $\nu = 4/3$ and fractal dimension $d_f = 91/48$. The situation is more complex for rigidity percolation (RP). While recent studies have investigated conformal invariance in 2D RP~\cite{Javerzat_2023} and provided numerical evidence that RP interfaces are consistent with a Schramm-Loewner evolution (SLE) description with $\kappa \approx 2.9$~\cite{javerzat_2024} (to be compared with $\kappa = 8/3$ for the external perimeter of critical BP), exact solutions for RP remain unavailable. Consequently, understanding RP relies heavily on numerical simulations. For bond RP on the triangular lattice, numerical studies yield a critical threshold $p_c = 0.6602(3)$~\cite{jacobs1996generic}. The correlation length exponent $\nu = 1.19(1)$~\cite{Javerzat_2023} describes how the correlation length $\xi$ diverges near the critical point as $\xi \sim |p-p_c|^{-\nu}$, characterizing the typical length scale over which fluctuations are correlated. The fractal dimension $d_f = 1.853(5)$~\cite{moukarzel1999comparison} characterizes the mass scaling of critical clusters, where the cluster size $S$ scales with the radius of gyration $R_s$ as $S \sim R_s^{d_f}$. These exponents differ from those of BP, confirming that the two models belong to different universality classes. Recent advances in understanding constrained systems~\cite{rivoire2006exactly,henkes2016rigid,goodrich2012finite} have deepened our understanding of rigidity-related phase transitions and further underscored the importance of characterizing RP and extracting its precise critical points and critical exponents. However, obtaining more accurate critical point and critical exponents remains challenging due to exceptionally strong finite-size corrections that persist even for systems with side length $L = 3200$, and estimates of critical exponents, particularly $d_f$, vary depending on the analysis method employed~\cite{moukarzel1999comparison}.

The distinction between rigidity and Bernoulli percolation lies in their underlying nature: while Bernoulli percolation involves scalar properties determined solely by topological adjacency, rigidity percolation requires the satisfaction of vectorial force balance equations~\cite{moukarzel1999comparison}. Beyond this vectorial character, rigidity percolation exhibits an inherent non-local aspect that is absent in BP~\cite{jacobs1996generic}. For instance, whether two rigid clusters are mutually rigid depends not only on their direct connections (shared sites) but also on the mechanical constraints imposed by distant clusters that share no direct connections with either of them. This non-local coupling is exemplified by the ``house-of-cards'' mechanism~\cite{moukarzel1999comparison}, where the removal of a single critical bond can break a large rigid cluster into numerous small rigid pieces. Conversely, the addition of a critical bond can trigger the simultaneous merger of many previously independent rigid clusters, demonstrating that such critical bonds can substantially alter the system's mechanical properties. However, conventional static ensembles, which measure observables after all bonds have been placed at fixed occupation probabilities or concentrations, are poorly suited to capture these cascade events. Static measurements cannot capture the changes before and after adding each bond, thereby potentially masking the dynamical mechanisms underlying the rigidity transition. This motivates the need for dynamic measurements capable of capturing behavior at specific moments during the bond-addition process.

Dynamic simulations that progressively add bonds to an initially empty lattice offer an alternative approach to studying RP. In such simulations, one starts from an empty lattice and sequentially adds bonds by randomly selecting an unoccupied bond at each step. We define a dimensionless ``time'' $t = T/E$, where $T$ is the number of bonds added and $E$ is the total number of bonds in the system. Here, $t$ serves as the temporal control parameter in dynamic simulations, corresponding to the bond occupation probability $p$ used in traditional static ensembles. Throughout this paper, we use $t_c$ and $p_c$ equivalently to denote the critical point. At each time step, rigid clusters are identified and quantities of interest, such as the largest cluster size, are measured and recorded~\cite{lu2024self}. Each individual realization of the dynamic process defines its own pseudo-critical point $t_L$ based on specific events during the bond-addition process, where $t_L$ is the value of $t$ at which the specific event occurs. This methodology enables the construction of event-based ensembles~\cite{li2023explosive}, where statistical averages are performed over configurations selected according to the occurrence of particular dynamical events rather than fixed parameter values. Early work by Moukarzel and Duxbury employed this methodology on body-bar lattices, identifying the moment when a system-spanning rigid cluster first appears as the sample-specific percolation threshold~\cite{moukarzel1995stressed}. Dynamic simulations enabled the study of ensemble properties at these dynamically defined critical points. In this framework, the pseudo-critical point becomes an observable rather than a control parameter, and its measurement is straightforward. One can study the finite-size scaling (FSS) of pseudo-critical points $t_L$ according to $t_L - t_c \sim L^{-1/\nu}$ and their fluctuations scaling as $\sigma_{t_L} = \sqrt{\langle t_L^2 \rangle - \langle t_L \rangle^2} \sim L^{-1/\nu}$. Additionally, one can characterize other observables such as the largest cluster size $C_1$ in these event-based ensembles. The versatility of this framework lies in the freedom to define pseudo-critical points through various dynamical events, allowing for different event-based ensembles that enable detailed examination of critical behavior near pseudo-critical point.

Recent advances in percolation studies have demonstrated the effectiveness of event-based ensembles in revealing clean critical behavior and precise critical exponents. In explosive percolation, traditional fixed-density ensembles exhibit anomalous scaling behavior, including non-self-averaging properties~\cite{riordan2012achlioptas}, whereas event-based approaches that identify pseudo-critical points through the largest gap in the growth of the largest cluster, $C_1$, have revealed clean critical behavior conforming to standard FSS theory~\cite{li2023explosive,li2024explosive}. Similarly, in high-dimensional percolation ($d > 6$) with open boundary conditions, conventional ensembles at fixed bond occupation probability $p_c$ display anomalous FSS behavior, with measured effective exponents deviating from theoretical predictions and exhibiting strong finite-size corrections. However, event-based ensembles at pseudo-critical points restore clean standard FSS behavior without such complications~\cite{li2024crossover}. These findings suggest that event-based ensembles may also provide clean scaling behavior and enable high-precision determination of critical properties in rigidity percolation.

However, a dynamic algorithm capable of tracking complete rigid cluster information at each bond addition step has not been readily available. The computational landscape of rigidity percolation has been defined by two established algorithms. The first is the Pebble Game (PG) algorithm by Jacobs and Hendrickson~\cite{jacobs1996generic}, which conceptualizes degrees of freedom as ``pebbles'' to identify rigid clusters and floppy modes but operates statically, determining rigidity only after all bonds have been placed. The second is Moukarzel's incremental procedure~\cite{moukarzel1995stressed}, which, while designed for dynamic simulations, employs a recursive ``body-bar'' condensation technique that only identifies over-constrained subgraphs upon detecting redundant bonds, rather than providing complete rigid cluster information at each step. The natural approach would be to adapt the PG algorithm for dynamic use. A dynamic implementation of the PG algorithm would necessarily involve cluster merger operations after each bond addition, but naive implementations of these operations incur substantial computational costs. Such operations require traversing all sites within merging clusters and their neighbors, and near the critical point and in super-critical regions, where rigid clusters become extensive, these frequent traversals of large clusters become computationally prohibitive.

To overcome this limitation, we introduce a dynamic PG algorithm that efficiently tracks the evolution of rigid clusters as bonds are added sequentially. By leveraging the current cluster information, our algorithm reduces the determination of bond independence to $O(1)$ time complexity. Our modified algorithm employs a ``let go of the largest'' technique during rigid cluster identification, achieving computational efficiency comparable to that of static PG algorithm implementations. With this efficient algorithm, we performed simulations on systems up to the linear side length $L = 8192$ with periodic boundary conditions, enabling a systematic analysis of cluster merger dynamics. This approach reveals several notable findings. First, we discover a previously unrecognized temporal self-similarity in rigidity percolation dynamics, where the temporal distributions of cluster size changes (gaps) and the number of clusters merged upon single bond additions both follow universal power laws. Second, we define event-based ensembles that yield high-precision estimates: the critical point $p_c = 0.660\,277\,8(10)$, representing an improvement of three orders of magnitude over the previous result $p_c = 0.6602(3)$ from Ref.~\cite{jacobs1996generic}, the inverse correlation-length exponent $1/\nu = 0.850(3)$, and the fractal dimension $d_f = 1.850(2)$.

Cascade merger events, in which a single bond triggers the simultaneous coalescence of many clusters, were identified in Ref.~\cite{moukarzel1999comparison} and proven rigorously for complete graphs~\cite{kasiviswanathan2011rigidity,fernholz2007k,cain2007random}, yet a quantitative characterization of their scaling properties has been lacking. To our knowledge, our simulations provide the first such characterization on the triangular lattice. We find that the maximum number of merged clusters scales with system size and the governing exponent is consistent with the inverse correlation-length exponent $1/\nu=0.850(3)$. Our analysis of these cascade merger events provides a heuristic explanation for the exceptionally strong finite-size corrections observed in rigidity percolation.

The remainder of this paper is organized as follows: Section~\ref{Sec:DynamicAlgorithm} presents our dynamic PG algorithm. Section~\ref{Sec:SampledObservables} defines the observables and event-based ensembles. Section~\ref{Sec:Results} analyzes the scaling of pseudo-critical points, cascade dynamics, static and dynamic cluster distributions, and the scaling analysis of quantities related to the largest cluster near pseudo-critical points. Finally, Section~\ref{Sec:Conclusion} summarizes our findings and outlines future research directions.

\section{Dynamic Algorithm}
\label{Sec:DynamicAlgorithm}

We employ a dynamic simulation where bonds are added sequentially to an empty triangular lattice. This approach requires an efficient method for tracking rigid clusters dynamically as the system evolves. The full details of our dynamic PG algorithm are presented in Appendix~\ref{app:dynamic_pebble_game}; here, we discuss the computational challenges and summarize the key features of our solution.

\subsection{Computational Challenges}

After each bond addition, we must identify all rigid clusters in the current configuration. This requires two essential tasks: (i) determining whether the newly added bond is independent (contributes to system rigidity), and (ii) if independent, determining whether the new bond induces mergers of other existing rigid clusters and identifying which clusters are merged.

The first task—determining bond independence—can be accomplished using the standard PG algorithm~\cite{jacobs1996generic}, and this application does not increase the time complexity compared to its use in conventional static simulations. However, the second task—merging the clusters that should be merged—poses a severe computational challenge. In BP, adding a bond can only merge the two connectivity clusters to which the bond's endpoints belong. In contrast, the non-local nature of rigidity constraints in RP means that adding a single bond can trigger the simultaneous merger of many clusters, including those that are not directly connected to the new bond. Checking only the neighboring clusters of the new bond is insufficient in RP.

\begin{figure}[tp]
    \centering
    \includegraphics[width=0.6\linewidth]{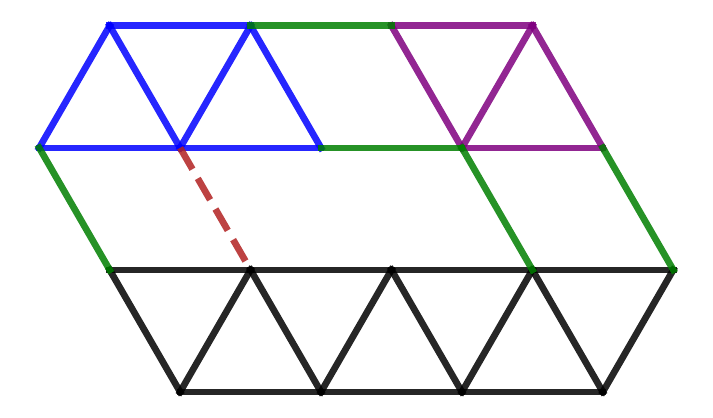}
    \caption{Illustration of a multi-cluster merger event in rigidity percolation on the triangular lattice. Different colors represent distinct rigid clusters before the addition of the new bond (shown as a dashed red line), with each green bond representing a single-bond rigid cluster consisting of one bond and two sites. Upon adding this bond, nine rigid clusters—including the rigid cluster formed by the new bond itself—merge into a single large rigid cluster. Notably, the merger involves clusters (such as the green and purple clusters) that are not directly connected to the newly added bond, exemplifying the non-local characteristic of rigidity percolation. To identify all merging clusters, one must traverse not only the clusters directly connected by the new bond (blue and black clusters), but also their neighboring clusters (green clusters), and iteratively check the neighbors of those clusters (purple cluster), adding newly discovered merging clusters to the traversal queue until no unvisited clusters remain.}
    \label{fig:nine_merge}
\end{figure}

Figure~\ref{fig:nine_merge} illustrates this complexity. The dashed red line represents the newly added bond, which is an independent bond. Different colors mark distinct rigid clusters before this bond addition, with each green bond representing a single-bond rigid cluster. Since the new bond is independent, it forms a rigid cluster by itself. If it merges with other clusters, then the new bond together with those clusters form a larger rigid cluster. In this example, the figure shows that nine clusters merge simultaneously. Crucially, only the blue and the black clusters are directly connected to the new bond. To detect the complete set of merging clusters, we must: (1) check whether the blue and black clusters merge; (2) examine their neighbors (the green clusters) to determine if they also merge; (3) iteratively check the neighbors of the green clusters (the purple cluster); and (4) continue this process, adding newly discovered merging clusters to the traversal queue, until no unvisited clusters remain. This procedure requires traversing all sites belonging to the merging clusters and examining their neighboring sites to identify potential mergers.

The computational cost becomes prohibitive near the critical point and in the supercritical region. When the largest rigid cluster reaches a size of $O(L^2)$, any merger event that incorporates the largest cluster into the merged cluster requires traversing all its sites. Since such events occur $O(L^2)$ times during the bond-addition process from empty to full occupation, the total computational complexity scales as $O(L^2 \times L^2) = O(L^4)$ for the supercritical phase alone. This is far worse than the $O(V^{1.15})$ complexity (where $V = L^2$ is the total number of vertices) achieved by the standard PG algorithm in conventional fixed-probability ensembles~\cite{jacobs1997algorithm}, rendering large-scale dynamic simulations impractical without algorithmic innovations.

\subsection{Our Dynamic Pebble Game Algorithm}

To overcome these limitations, we have developed an efficient dynamic PG algorithm with two key innovations. First, we maintain persistent data structures that store cluster membership information at each time step, including bidirectional mappings between vertices and the clusters to which they belong. This enables the classification of bond independence in $O(1)$ time without requiring pebble searches, reducing computational overhead compared to repeated applications of the standard PG algorithm.

Second, we employ a ``let go of the largest'' optimization during cluster merger identification (detailed in Appendix~\ref{app:dynamic_pebble_game}). Instead of traversing all merging clusters, we exclude the largest cluster among the merging clusters from the traversal process. This substantially reduces the number of site visits, particularly near the critical point and in the supercritical region. Combined with efficient data structures, our algorithm achieves an overall computational complexity of $O(V^{1.15})$ shown in Fig.~\ref{fig:performance}, matching the efficiency of static methods~\cite{jacobs1997algorithm}.

During the preparation of this manuscript, a similar dynamic algorithm was proposed in a recent preprint~\cite{javerzat2025fast}, which cites an earlier, short article of ours where the algorithm was not detailed. The main algorithmic optimizations in both approaches are similar, with the key differences lying in the data structure implementations. While the algorithm in Ref.~\cite{javerzat2025fast} has a reported scaling of $O(V^{1.05})$, our implementation is found to be faster compared to their code on Github~\footnote{The algorithm in Ref.~\cite{javerzat2025fast} includes an extra optimization that uses connectivity cluster properties to determine whether a merger check is necessary after bond addition. This provides a slight improvement but does not change the overall time complexity scaling. Our implementation, which uses different data structures, is found to be faster in practice; for example, for a system of size $L = 2048$, our algorithm is approximately ten times faster when tested on an Intel(R) Core(R) E5-2680 v4 @ 2.40GHz.}.

\section{Sampled Observables}
\label{Sec:SampledObservables}

We simulate bond rigidity percolation on triangular lattices with periodic boundary conditions and side length $L$, where the total number of bonds is $E=3L^2$. To better resolve finite-size corrections, we simulate a comprehensive range of system sizes. In addition to standard powers of two ($L = 4, 8, \ldots, 8192$), we also simulate intermediate sizes (e.g., $L = 320, 768, 1536, 3072$). For each system size, we generate on the order of $10^6$ independent samples (e.g., $2.4 \times 10^6$ samples for $L = 2048$), except for $L = 8192$, which has $3 \times 10^5$ samples. For the bond configuration at each step $T$, we record the following dynamic quantities:
\begin{enumerate}
    \item The size of each rigid cluster, $\mathcal{C}(T)$, defined as the number of vertices in the cluster.
    \item The size of the largest rigid cluster, $\mathcal{C}_1(T)$.
    \item The number of clusters merged, $\mathcal{K}(T)$. When a new bond is added at step $T$, it might merge $\mathcal{K}(T) \ge 3$ clusters that were not previously mutually rigid. If no merger occurs, we set $\mathcal{K}(T)=0$.
    \item The merged cluster size, $\mathcal{S}(T)$. If adding a bond at step $T$ results in a merger of $\mathcal{K}(T)$ clusters, the merged cluster has size $\mathcal{S}(T)$, which is the sum of all participating cluster sizes minus the contributions from shared vertices. If no merger occurs, there are two cases: if the new bond is independent, $\mathcal{S}(T) = 2$; if the new bond is redundant, $\mathcal{S}(T)$ equals the size of the cluster to which the new bond belongs.
    \item The gap, $\mathcal{G}(T)$. If adding a bond at step $T$ results in a merger of $\mathcal{K}(T)$ clusters with sizes $s_1, \dots, s_{\mathcal{K}(T)}$, the gap is defined as
    \[ \mathcal{G}(T) = \mathcal{S}(T) - \max_{i} \{s_i\}. \]
    If no merger occurs, we set $\mathcal{G}(T)=0$.
    \item The susceptibility $\chi(T) = L^{-d} \langle \sum_{\mathcal{C} \neq \mathcal{C}_1} |\mathcal{C}|^2 \rangle$, where the summation excludes the largest cluster. This quantity is also referred to as the second moment of the cluster-size distribution.
\end{enumerate}

To illustrate these quantities with a concrete example, consider the nine-cluster merger event shown in Fig.~\ref{fig:nine_merge}. In this configuration, adding the dashed red bond triggers the simultaneous merger of nine rigid clusters ($\mathcal{K} = 9$). The merged cluster has size $\mathcal{S} = 18$. The largest cluster before the merger is the black cluster with size 9. Therefore, the gap is $\mathcal{G} = \mathcal{S} - 9 = 9$.

These dynamic quantities exhibit distinct behaviors across different stages of the bond-addition process. At early times, bond additions rarely trigger mergers; a typical rigid cluster has size $O(1)$, and both $\mathcal{C}_1$ and $\chi$ remain small. As more bonds are added, merger events become gradually more frequent, so the averages of $\mathcal{K}$ and $\mathcal{G}$ increase, and $\mathcal{C}_1$ and $\chi$ grow steadily. Near the critical point, $\mathcal{K}$ and $\mathcal{G}$ reach their peak values, signaling rapid large-scale coalescence; $\chi$ also peaks, indicating that clusters other than $\mathcal{C}_1$ have grown large. The subsequent merger of these large clusters with $\mathcal{C}_1$ produces a large gap and drives substantial growth of the largest cluster. Beyond the critical region, merger activity subsides: $\mathcal{K}$, $\mathcal{G}$, and $\chi$ decrease, while $\mathcal{C}_1$ continues to grow and spans over the entire system. From the evolution of these quantities, we identify three definitions for the pseudo-critical points of each dynamic process:
\begin{enumerate}[label=(\alph*)]
    \item The gap pseudo-critical point, $\mathcal{T}_G$, is the time when the gap reaches its maximum.
    \item The susceptibility pseudo-critical point, $\mathcal{T}_{\chi}$, is the time when the susceptibility $\chi(T)$ reaches its maximum.
    \item The merger pseudo-critical point, $\mathcal{T}_K$, is the time when the maximum number of clusters merge simultaneously, i.e., when $\mathcal{K}(T)$ is maximized.
\end{enumerate}

The collection of system configurations at which these pseudo-critical points are identified constitute three distinct event-based ensembles, indexed by $\alpha \in \{G, \chi, K\}$. We then compute the following quantities:
\begin{enumerate}[label=(\roman*)]
    \item The average pseudo-critical points $t_{\alpha} = \langle \mathcal{T}_{\alpha} \rangle$ and their standard deviations $\sigma_{t_{\alpha}} = \sqrt{\langle \mathcal{T}_{\alpha}^2 \rangle - \langle \mathcal{T}_{\alpha} \rangle^2}$.
    \item The average size of the largest cluster $C_{1,t_\alpha} = \langle \mathcal{C}_1 \rangle_{t_\alpha}$.
    \item The average number of merged clusters $K_{t_\alpha} = \langle \mathcal{K} \rangle_{t_\alpha}$.
\end{enumerate}

In addition to the event-based ensembles defined above, we also measure probability distributions in both static and dynamic contexts. The static cluster size distribution, denoted $P_s(s)$, is measured on a single snapshot of the system configuration—either at a fixed bond occupation probability $p$ or at a pseudo-critical point. It quantifies the number of clusters of size $s$ divided by the system volume $V$. In contrast, we define dynamic probability density functions $\mathcal{P}_S(s)$ and $\mathcal{P}_G(s)$ that aggregate statistics over the entire bond-addition process from $t=0$ to a specified termination time $t_e$. Here, $\mathcal{P}_S(s)$ counts the number of merger events that produce a merged cluster of size $s$ (i.e., $\mathcal{S}(T) = s$), normalized by the system volume $V$. Similarly, $\mathcal{P}_G(s)$ counts the number of merger events that produce a gap of size $s$ (i.e., $\mathcal{G}(T) = s$), also normalized by $V$. These dynamic distributions capture the temporal statistics of cluster mergers throughout the bond-addition process.

\section{Results}
\label{Sec:Results}

We perform in this section a detailed data analysis to various quantities. To determine the critical exponents and other universal quantities from our numerical data, we employ a least-squares fitting procedure. For a generic observable $\mathcal{O}$, the data are fitted to the expected scaling form, which typically includes corrections to scaling:
\begin{equation}
    \mathcal{O}(L) = L^{y_{\mathcal{O}}} (a_0 + a_1 L^{-y_1} + a_2 L^{-y_2}) + c_0,
    \label{eq:O_fit}
\end{equation}
where $y_{\mathcal{O}}$ is the leading scaling exponent, $y_1, y_2 > 0$ are correction exponents, and $c_0$ is a non-universal constant.

To mitigate the influence of higher-order corrections not explicitly included in our fitting function, we introduce a lower cutoff, $L_{\min}$, such that only data for system sizes $L \ge L_{\min}$ are considered. We then systematically increase $L_{\min}$ to test the stability of the results. A fit is deemed reliable if it satisfies two conditions: the goodness-of-fit, quantified by $\chi^2$ per degree of freedom (DF), should be of order unity ($\chi^2/\text{DF} \approx 1$), and the resulting parameter estimates must be stable against further increases in $L_{\min}$. The final systematic uncertainties are estimated by analyzing the variation in the fitted parameters when using different, but equally plausible, fitting ansatz.

\subsection{Pseudo-critical points}
\label{subsec:pseudo_critical_points}

We first analyze the FSS behavior of the pseudo-critical points and their fluctuations. According to FSS theory, the average pseudo-critical point $t_{\alpha}(L) = \langle \mathcal{T}_{\alpha}(L) \rangle$ for a system of linear size $L$ approaches the critical point $t_c$ as
\begin{equation}
    t_{\alpha} - t_c \sim L^{-1/\nu},
    \label{eq:pseudo_critical_scaling}
\end{equation}
and its fluctuations are expected to scale as
\begin{equation}
    \sigma_{t_{\alpha}} = \sqrt{\langle \mathcal{T}_{\alpha}^2 \rangle - t_{\alpha}^2} \sim L^{-1/\nu},
    \label{eq:pseudo_critical_fluctuation}
\end{equation}
where $\nu$ is the correlation-length exponent.

\begin{figure}[tp]
    \centering
    \includegraphics[width=\linewidth]{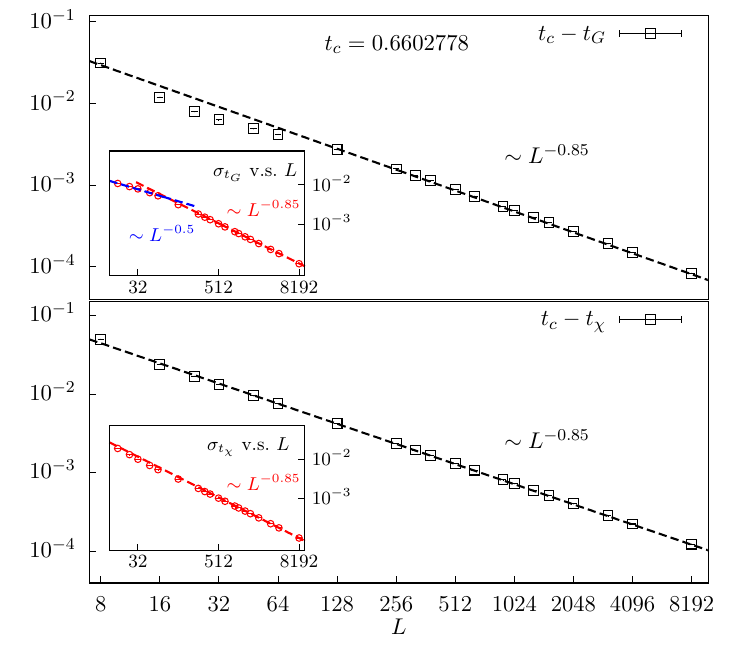}
    \caption{Finite-size scaling of the pseudo-critical points $t_G$ (top) and $t_{\chi}$ (bottom). Main panels show $t_c - t_{\alpha}$ vs. $L$ on a log-log scale, with $t_c = 0.660\,277\,8$. Dashed lines represent a power-law behavior $\sim L^{-0.85}$. Insets show the fluctuations of pseudo-critical points $\sigma_{t_{\alpha}}$ vs. $L$. The dashed lines in the insets for large $L$ correspond to $\sim L^{-0.85}$. For $L < 128$, the fluctuation $\sigma_{t_G}$ is well described by $\sim L^{-0.5}$ (blue dashed line).}
    \label{fig:pseudo_points}
\end{figure}

Figure~\ref{fig:pseudo_points} presents the scaling behavior for $t_G$ and $t_{\chi}$. The value, $0.660\,277\,8$, of $t_c$ used in the plots was determined from a subsequent numerical analysis of the pseudo-critical point, $t_{\chi}$, whose finite-size corrections are easier to handle than those of $t_G$. The data show that both pseudo-critical points and their fluctuations follow a power-law scaling with system size, which is consistent with Eqs.~\eqref{eq:pseudo_critical_scaling} and \eqref{eq:pseudo_critical_fluctuation}. Although $t_G$ and $t_{\chi}$ are defined from different dynamical events, the data indicate that they converge to the same critical point $t_c$ and are governed by the same scaling exponent in the large-$L$ limit.

As can also be seen in Fig.~\ref{fig:pseudo_points}, for small systems, $t_G$ deviates more from the scaling $\sim L^{-0.85}$ and exhibits non-monotonic asymptotic behavior, indicating the presence of complex finite-size corrections. In contrast, $t_{\chi}$ closely follows the scaling $\sim L^{-0.85}$ even at small scales, suggesting only minor finite-size corrections. Therefore, $t_{\chi}$ is a more suitable quantity for high-precision estimation of the critical parameters.

To obtain precise estimates, we perform detailed FSS analysis for both the fluctuation $\sigma_{t_{\chi}}$ and the pseudo-critical point $t_{\chi}$ itself using the FSS form in Eq.~\eqref{eq:O_fit}. For the pseudo-critical points, we set $y_{\mathcal{O}} = -1/\nu$ and $c_0 = t_c$. For the fluctuations, we set $y_{\mathcal{O}} = -1/\nu$ and $c_0 = 0$. The fitting results are summarized in Table~\ref{tab:tchi_fits}. For the fluctuation $\sigma_{t_{\chi}}$, we first perform fits including correction terms with exponents $y_1 = -1$ and $y_2 = -2$. While these fits yield stable results, the large uncertainty in the coefficient $a_2$ indicates that the $-2$ term is redundant. This observation is consistent with the monotonic approach to the asymptotic behavior observed in Fig.~\ref{fig:pseudo_points}, suggesting that a single correction term is sufficient. We therefore perform additional fits including only the $y_1 = -1$ correction term, which also produce stable results. Our final estimate is $1/\nu = 0.850(3)$, which is consistent with the value $0.84(1)$ reported in Ref.~\cite{Javerzat_2023}.

We then fit the pseudo-critical point $t_{\chi}$ itself. When we include correction terms with exponents $y_1 = -1$ and $y_2 = -2$ and allow both $t_c$ and $1/\nu$ to vary, the fits yield stable results. As shown in Table~\ref{tab:tchi_fits}, the fitted value of $1/\nu$ is $0.855(5)$, which agrees with the exponent obtained from $\sigma_{t_{\chi}}$ within error bars. This consistency confirms that both quantities are governed by the same correlation length exponent, and we therefore adopt $1/\nu = 0.850(3)$ from the fluctuation analysis for subsequent fits. With this exponent fixed, we perform fits of $t_{\chi}$ with correction term $y_1 = -1$. The results are insensitive to the specific choice of correction exponents (not shown in the table). Given the excellent power-law behavior observed in the main panel of Fig.~\ref{fig:pseudo_points} and the consistency of the fitting results, we obtain our final estimate $t_c = 0.660\,277\,8(10)$, which improves the precision by three orders of magnitude compared to the previous most precise result of $0.6602(3)$~\cite{jacobs1996generic}.

\begin{table*}[htbp]
    \centering
    \caption{Fitting results for the fluctuation $\sigma_{t_{\chi}}$ and the pseudo-critical point $t_{\chi}$ using Eq.~\eqref{eq:O_fit}. For $\sigma_{t_{\chi}}$, we set $y_{\mathcal{O}} = -1/\nu$ and $c_0 = 0$. For $t_{\chi}$, we set $y_{\mathcal{O}} = -1/\nu$ and $c_0 = t_c$. For each fit, $L_{\mathrm{min}}$ indicates the minimum system size used. Values without error bars are fixed during the fitting procedure. A dash ($-$) indicates that the corresponding term was excluded from the fit. The final estimates are $1/\nu = 0.850(3)$ and $t_c = 0.660\,277\,8(10)$.}
    \label{tab:tchi_fits}
    \begin{threeparttable}
        \begin{tabular}{c l l l l l l l l l}
            \hline\hline
             & \multicolumn{1}{c}{$L_{\mathrm{min}}$} & \multicolumn{1}{c}{$t_c$} & \multicolumn{1}{c}{$-1/\nu$} & \multicolumn{1}{c}{$a_0$} & \multicolumn{1}{c}{$y_1$} & \multicolumn{1}{c}{$a_1$} & \multicolumn{1}{c}{$y_2$} & \multicolumn{1}{c}{$a_2$} & \multicolumn{1}{c}{$\chi^2/\mathrm{DF}$} \\
            \hline
            \multirow{4}{*}{$\sigma_{t_{\chi}}$} & 384 & \multicolumn{1}{c}{$-$} & $-0.849(1)$ & 0.220(2) & $-1$ & $-6.7(6)$ & $-2$ & $715\pm133$ & 9.90/7 \\
             & 640 & \multicolumn{1}{c}{$-$} & $-0.852(2)$ & 0.225(4) & $-1$ & $-9\pm1$ & $-2$ & $1299\pm479$ & 7.18/5 \\
             & 896 & \multicolumn{1}{c}{$-$} & $-0.848(1)$ & 0.218(2) & $-1$ & $-5.3(5)$ & \multicolumn{1}{c}{$-$} & \multicolumn{1}{c}{$-$} & 11.47/5 \\
             & 1024 & \multicolumn{1}{c}{$-$} & $-0.850(1)$ & 0.220(3) & $-1$ & $-6.1(6)$ & \multicolumn{1}{c}{$-$} & \multicolumn{1}{c}{$-$} & 7.07/4 \\
            \hline
            \multirow{4}{*}{$t_{\chi}$} & 512 & 0.660274(1) & $-0.870(9)$ & $-0.31(2)$ & $-1$ & $8\pm4$ & $-2$ & $1488\pm935$ & 2.32/5 \\
             & 512 & 0.660276(1) & $-0.856(3)$ & $-0.276(6)$ & $-1$ & $1.3(5)$ & \multicolumn{1}{c}{$-$} & \multicolumn{1}{c}{$-$} & 5.35/6 \\
             & 384 & 0.6602780(2) & $-0.85$ & $-0.2645(2)$ & $-1$ & $0.33(8)$ & \multicolumn{1}{c}{$-$} & \multicolumn{1}{c}{$-$} & 11.43/8 \\
             & 640 & 0.6602778(2) & $-0.85$ & $-0.2643(2)$ & $-1$ & $0.2(2)$ & \multicolumn{1}{c}{$-$} & \multicolumn{1}{c}{$-$} & 9.79/6 \\
            \hline\hline
        \end{tabular}
    \end{threeparttable}
\end{table*}

Intriguingly, the fluctuation $\sigma_{t_G}$ exhibits complex finite-size corrections. For small system sizes ($L < 128$), the data align closely with the scaling $\sigma_{t_G} \sim L^{-0.5}$, which is markedly different from the asymptotic behavior $\sigma_{t_G} \sim L^{-0.85}$ seen for larger $L$. Notably, the effective exponent of $-0.5$ is even larger than the value $-1/\nu \approx -0.75$ for 2D BP. This behavior can be understood as a result of two competing effects. On one hand, merging clusters in rigidity percolation is more difficult than in 2D BP, as it requires establishing mutual rigidity rather than simple connectivity. This tends to broaden the critical region. On the other hand, the addition of a single bond can trigger a cascade that merges an extensive number of previously independent clusters, an effect that would narrow the transition. At small system sizes, geometric constraints may limit the availability of patterns for these large-scale cascades. In this regime, the difficulty of merging dominates, leading to a wider effective critical region and an exponent that is larger than that of 2D BP. As $L$ increases, large-scale merger events become more prevalent, and the system approaches its true asymptotic behavior, which is characterized by a narrower critical window. This highlights a practical challenge for numerical studies: for some quantities, the finite-size corrections are so pronounced that only data from very large systems can yield reliable estimates of the asymptotic exponents.

\begin{figure}[tp]
    \centering
    \includegraphics[width=\linewidth]{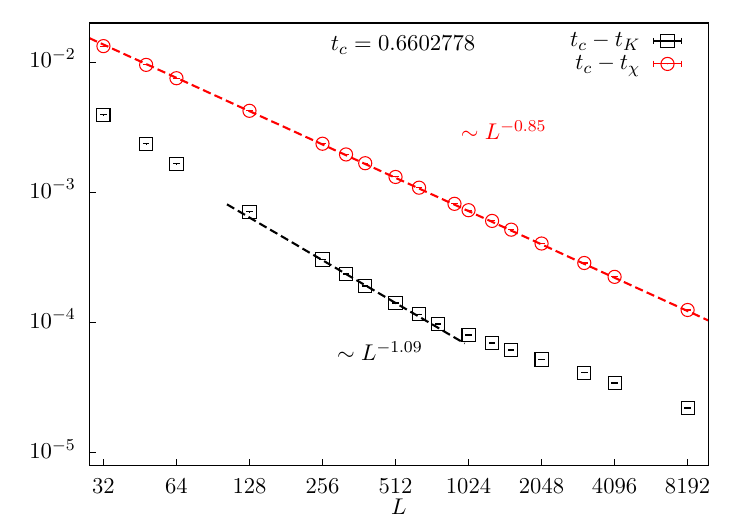}
    \caption{Comparison of the finite-size scaling behavior for pseudo-critical points $t_K$ (black squares) and $t_{\chi}$ (red circles). The plot shows $t_c - t_K$ and $t_c - t_{\chi}$ versus system size $L$ on a log-log scale. The dashed lines indicate power laws with exponents $-1.09$ and $-0.85$, respectively. For small to moderate system sizes, $t_K$ appears to converge faster, but this reflects the dominance of a subleading correction term rather than the true critical scaling. Notably, the data demonstrate that $t_K$ at $L=512$ is even closer to $t_c$ than $t_{\chi}$ at $L=8192$.}
    \label{fig:tK_compare}
\end{figure}

We further define a pseudo-critical point $t_K$ as the time when the maximum number of clusters merge simultaneously. Remarkably, this quantity approaches $t_c$ exceptionally fast. As shown in Fig.~\ref{fig:tK_compare}, the finite-size value $t_K$ for $L=512$ is already closer to $t_c$ than $t_{\chi}$ for $L=8192$. The data in the main panel reveal that for small to moderate system sizes, $t_c - t_K$ appears to follow a power law with an exponent of approximately $-1.09$, which is substantially steeper than the $-0.85$ scaling observed for $t_{\chi}$. However, this behavior is deceptive. The true leading scaling term of $t_K$ shares the same critical exponent as $t_{\chi}$, namely $-1/\nu \approx -0.85$, but with a very small coefficient. At small system sizes, this critical term is masked by a subleading correction term with exponent approximately $-1.09$. As the system size increases, the competition between these two terms leads to complex finite-size behavior, and while $t_K$ can still be fitted, this competing effect makes such fits more challenging and less reliable. For this reason, we use $t_{\chi}$ rather than $t_K$ for our determination of $t_c$. This further illustrates the exceptionally strong finite-size corrections in RP: without data from very large systems~($L = 4096$ and $8192$) and the high-precision $t_c$ from $t_{\chi}$, the apparent $-1.09$ scaling could misleadingly be interpreted as the leading critical exponent, leading to an incorrect determination of $t_c$.

\subsection{Cascade Dynamics}
\label{subsec:Cascade Dynamics}

As illustrated in Fig.~\ref{fig:nine_merge}, multiple rigid clusters can merge simultaneously in RP; we now investigate the underlying mechanisms and numerically characterize the behavior of large-scale merger events. The key lies in the rules governing cluster mergers. In BP, adding a bond can merge at most two connectivity clusters—simply connecting them is sufficient. In RP, mergers require not only connectivity but also the satisfaction of rigidity constraints: the merged structure must have no internal floppy modes except the trivial rigid-body motions. Consequently, even if multiple clusters belong to the same connectivity cluster, they may still remain distinct rigid clusters. However, a critical bond can trigger an avalanche-like merger that unifies an extensive number of such previously independent rigid clusters within the same connectivity cluster into a single rigid cluster. This phenomenon arises from two intertwined mechanisms: elementary multi-cluster merger patterns and a cascade effect.

\begin{figure}[tp]
    \centering
    \includegraphics[width=\linewidth]{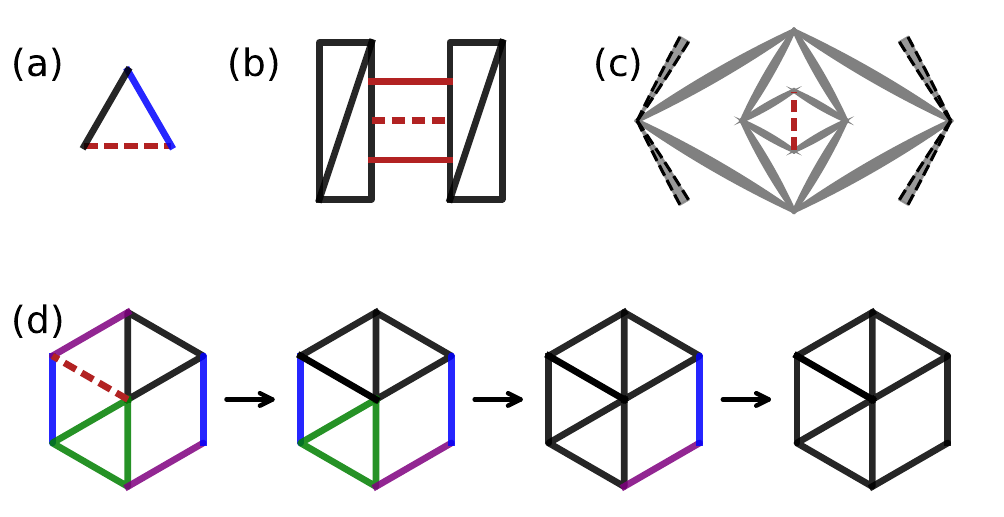}
    \caption{Examples of cluster merging patterns and cascade dynamics. Red dashed lines indicate newly added bonds that trigger the merging process. (a) The elementary \textit{triadic merger}, where three clusters merge simultaneously. (b) A five-cluster merger motif demonstrating more complex irreducible merger patterns. (c) An eye-shaped demo illustrating the cascade effect, where the addition of a single bond (red dashed line) can initiate a continuous sequence of triadic mergers. (d) A detailed cascade process showing how seven initially independent clusters (including the newly added bond as a rigid cluster) progressively merge through a series of triadic mergers, ultimately forming a single rigid body.}
    \label{fig:cascade}
\end{figure}

First, there are irreducible patterns by which multiple clusters can merge simultaneously. These patterns are ``irreducible'' in the sense that they cannot be decomposed into a sequence of mergers involving fewer clusters. It can be shown (Appendix.~\ref{app:odd_merger_proof}) that any merger event must involve an odd number of clusters. The simplest case, which we term a \textit{triadic merger}, involves three clusters merging, where each pair shares a common pivot, as depicted in Fig.~\ref{fig:cascade}(a). Before adding the red dashed bond, all vertices belong to the same connectivity cluster but are divided into two distinct rigid clusters. Adding this bond, which itself forms an independent rigid cluster, triggers the merger of these three rigid clusters into a single rigid cluster. A more complex example is the five-cluster merger shown in Fig.~\ref{fig:cascade}(b), where two of the clusters each share a pivot with the other three, but these two clusters do not share a pivot with each other, nor do the other three share pivots among themselves. Figure~\ref{fig:nine_merge} provides another example of an irreducible merger pattern involving nine rigid clusters. This merger is also irreducible because no subset of fewer than nine clusters can form a rigid cluster—all nine are essential to establish rigidity. Second, the formation of a new, larger rigid cluster via one of these elementary mergers can, in turn, enable it to merge with other neighboring clusters. This leads to a cascade effect, as illustrated in Fig.~\ref{fig:cascade}(c), where the eye-shaped configuration demonstrates how the addition of a single bond (red dashed line) can trigger a continuous and in principle infinite sequence of triadic mergers. A concrete cascade process is shown in Fig.~\ref{fig:cascade}(d), where the addition of a single bond (red dashed line) initiates a sequence of triadic mergers that ultimately unifies seven distinct rigid bodies.

\begin{figure}[tp]
    \centering
    \includegraphics[width=\linewidth]{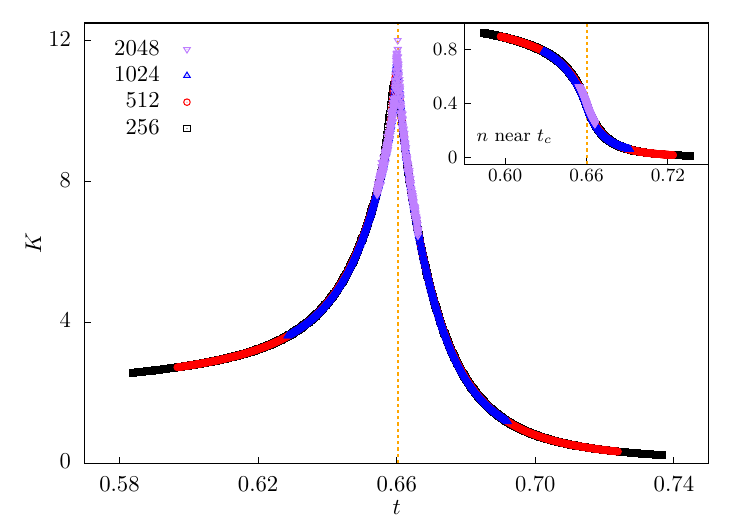}
    \caption{The mean number of merged clusters, $K(t) = \langle \mathcal{K}(t) \rangle$, as a function of time step $t$ for various system sizes $L$. The data for different $L$ collapse well, exhibiting a sharp peak at the critical point $t_c = 0.660\,277\,8(10)$, indicated by the dashed orange line. The inset shows the total number of clusters per site, $n$, versus $t$. This quantity decreases monotonically, with its fastest decline occurring around $t_c$ and showing minimal finite-size effects.}
    \label{fig:K_and_n}
\end{figure}

These two mechanisms together lead to large-scale merger events in RP. To quantify the number of merged clusters numerically, we employ our dynamic PG algorithm to measure the number of clusters that merge at each time step $t$, a quantity inaccessible in static simulations. Figure~\ref{fig:K_and_n} presents the average number of merged clusters $K(t) = \langle \mathcal{K}(t) \rangle$ and the total number of clusters per site $n(t)$. The data for $K(t)$ from different system sizes $L$ collapse onto a single curve that rises to a sharp peak near the critical point $t_c$ before falling. This demonstrates the critical nature of the large-scale merger events. The inset shows that the total number of clusters per site, $n(t)$, decreases monotonically. The rate of this decrease is maximal at $t_c$, consistent with the peak in merger activity observed in the main panel.

\begin{figure}[tp]
    \centering
    \includegraphics[width=\linewidth]{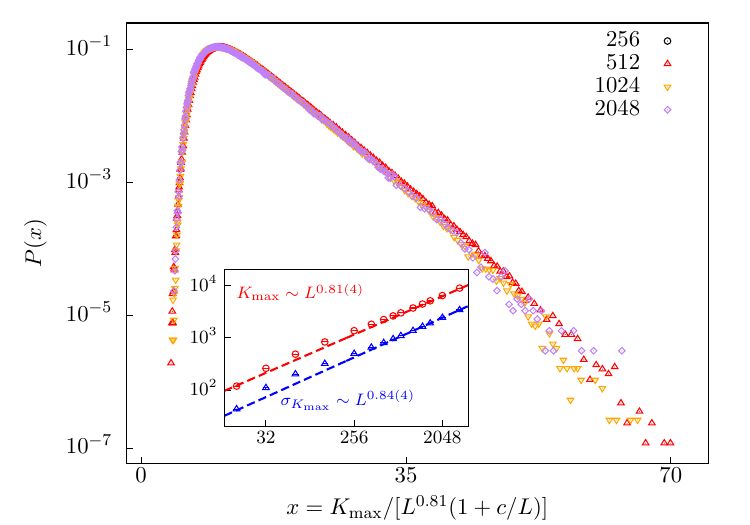}
    \caption{Scaling of the maximum number of merged clusters, $K_{\text{max}}$. The main panel shows the probability distribution $P(x)$ of the rescaled variable $x = K_{\text{max}}/[L^{0.81}(1+c/L)]$ with $c=158$. The data for different system sizes collapse onto a universal curve, which seems to approximately follow an exponential form as $e^{-bx}$ in the $x \to \infty$ limit, where $b$ is a non-universal constant. The inset shows the finite-size scaling of the average $\langle K_{\text{max}} \rangle$ and its fluctuation $\sigma_{K_{\text{max}}}$. The dashed lines are fits to the data, yielding scaling exponents of $0.81(4)$ and $0.84(4)$, respectively, consistent with the inverse correlation-length exponent $1/\nu=0.850(3)$.}
    \label{fig:K_max_dist}
\end{figure}

While the average $\langle \mathcal{K}(t) \rangle$ remains finite, the dynamics contain rare but extreme events. We probe these by measuring $\mathcal{K}_{\text{max}}$, the maximum number of clusters merged by a single bond within a stochastic dynamic process. As shown in the inset of Fig.~\ref{fig:K_max_dist}, its ensemble average, $K_{\text{max}} = \langle \mathcal{K}_{\text{max}} \rangle$, and fluctuation, $\sigma_{K_{\text{max}}} = \sqrt{\langle \mathcal{K}_{\text{max}}^2 \rangle - \langle \mathcal{K}_{\text{max}} \rangle^2}$, both diverge with system size $L$. Although the data scales well, the visible curvature indicates strong finite-size corrections. A fit to FSS ansatz Eq.~\eqref{eq:O_fit} yields scaling exponents of $0.81(4)$ for $K_{\text{max}}$ and $0.84(4)$ for $\sigma_{K_{\text{max}}}$. Notably, both exponents are consistent with the correlation length exponent $1/\nu = 0.850(3)$ determined from the pseudo-critical point analysis.

To further confirm this scaling, we analyze the probability distribution $P(x)$ of the rescaled variable $x = K_{\text{max}}/[L^{0.81}(1+c/L)]$, where the term with $c=158$ accounts for the leading finite-size correction. As shown in the main panel of Fig.~\ref{fig:K_max_dist}, the distributions for different system sizes collapse onto a single, universal curve with an approximately exponential tail at large $x$, which provides strong evidence for the well-defined scaling of $K_{\text{max}}$.

The observed large-scale merger events near $t_c$ provide physical insight into the substantial finite-size corrections in RP. To characterize the correlation length, we consider the spatial range spanned by the clusters that form the gap in a merger event, which reflects the extent over which a single bond addition alters the cluster membership of sites. In BP, where adding a bond merges at most two clusters, the correlation length $\xi_{\text{bp}}$ can be approximated by the gyration radius of the second-largest cluster, $R_{\text{bp}}$, throughout the bond-adding process. In RP, by contrast, the number of clusters forming the gap can vary widely. It may involve a small number of clusters with spatial extent on the order of a typical rigid cluster's gyration radius, $R_{\text{rp}}$, or it may involve extensive clusters distributed across an entire connectivity cluster with size $R_{\text{bp}}$, where $R_{\text{rp}} < R_{\text{bp}}$. Our numerical results support this picture: as shown in Fig.~\ref{fig:K_and_n}, intense merger activity occurs near $t_c$, and $K_{\text{max}}$ scales as $L^{0.81(4)}$, indicating that the number of merged clusters varies over a broad range. The presence of merger events with spatial extents ranging from $R_{\text{rp}}$ to $R_{\text{bp}}$ necessitates multiple correction terms to describe the finite-size behavior of cluster properties. Since both $R_{\text{rp}}$ and $R_{\text{bp}}$ grow with system size, they hinder convergence to the asymptotic scaling regime, thereby accounting for the persistence of finite-size corrections in RP.

\subsection{Anomalous Static Distributions}
\label{subsec:Anomalous Static Distributions}

The static cluster-size distribution $P_s$ represents a spatial snapshot that captures the statistical prevalence of all clusters across the entire lattice at a single time instant. It characterizes the critical phase through its power-law behavior. Near criticality, it is expected to follow the form $P_s \sim s^{-\tau} \tilde{\Phi}(s/L^{d_f})$, where $\tilde{\Phi}(x)$ is a universal cutoff function, and the Fisher exponent $\tau$ is given by the hyperscaling relation $\tau = 1 + d/d_f$. For 2D rigidity percolation ($d=2$, $d_f \approx 1.850$, see later in Section~\ref{subsec:df_difficulty}), this predicts $\tau \approx 2.08$.

\begin{figure}[tp]
    \centering
    \includegraphics[width=\linewidth]{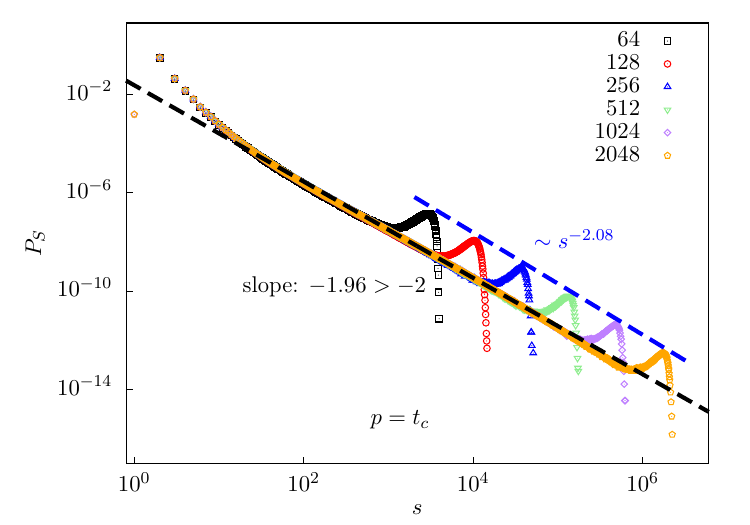}
    \caption{Static cluster size distribution $P_S$ versus cluster size $s$ in the traditional fixed bond probability ensemble at the critical point $t_c$ for various system sizes $L$. The power-law regime for intermediate cluster sizes exhibits an anomalous exponent of approximately $-1.96$ (black dashed line), significantly different from the theoretical prediction $-\tau \approx -2.08$ and even greater than $-2$. The blue dashed line with slope $-2.08$ shows the theoretically expected behavior for the characteristic cutoff region (hump), which aligns well with the data.}
    \label{fig:static_ns_pc}
\end{figure}

\begin{figure}[tp]
    \centering
    \includegraphics[width=\linewidth]{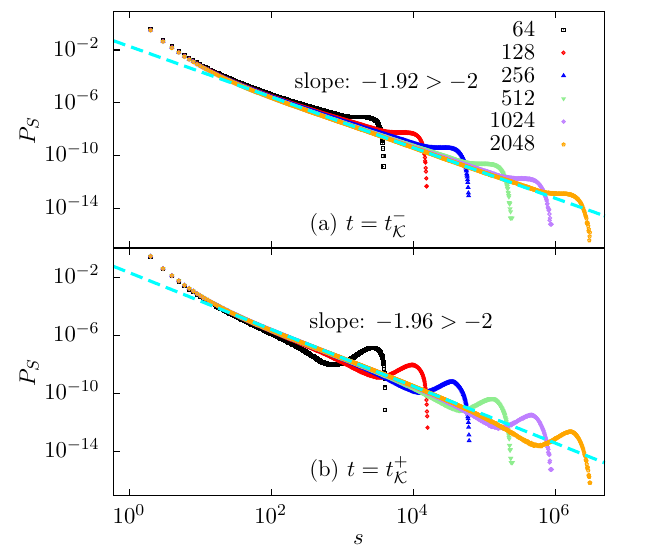}
    \caption{Static cluster size distribution $P_S$ versus cluster size $s$ in event-based ensembles defined by the moment when the maximum number of clusters merge simultaneously. (a) Distribution immediately before this maximum merger event ($t_K^-$) with anomalous exponent roughly -1.92. (b) Distribution immediately after this maximum merger event ($t_K^+$) with anomalous exponent around -1.96. The comparison reveals how the merger event reshapes the cluster distribution, depleting intermediate-sized clusters and forming the characteristic dip before the hump. }
    \label{fig:static_ns_tK}
\end{figure}

We investigate the static cluster distribution at the critical point in the traditional fixed bond probability ensemble, which provides a snapshot of all cluster sizes at a single, specific instant. Figure~\ref{fig:static_ns_pc} shows the cluster size distribution for various system sizes. The power-law regime for intermediate cluster sizes exhibits an anomalous exponent of approximately $-1.96$, which is markedly different from the theoretically predicted $-\tau \approx -2.08$. This anomalous exponent is even greater than $-2$, the threshold required for the normalizability of $P_s$ (ensuring $\sum s P_s$ converges). The anomalous scaling of intermediate clusters contrasts with the behavior of clusters near the characteristic cutoff size (the hump), which follows the theoretically expected slope of $-2.08$, as indicated by the blue dashed line in Fig.~\ref{fig:static_ns_pc}.

To further examine this anomalous crossover behavior, we analyze the cluster size distributions in event-based ensembles defined relative to the moment when the maximum number of clusters merge simultaneously. Figure~\ref{fig:static_ns_tK} presents distributions immediately before ($t_K^-$) and after ($t_K^+$) this maximum merger event in the upper and lower panels, respectively. This presentation clearly illustrates the transformation of the distribution across the merger event. Before the merger at $t_K^-$, the power-law exponent is approximately $-1.92$, while after the merger at $t_K^+$, it shifts to $-1.96$, closer to but still significantly different from the theoretical prediction. In both distributions, the tail region near the cutoff follows the expected slope of $-2.08$ like the distribution at $p = t_c$ (not shown in Fig.~\ref{fig:static_ns_tK}). 

The evolution across the maximum merger event provides further insight into this anomalous behavior. At $t_K^+$, a distinct dip forms before the hump, signifying the depletion of intermediate-sized clusters as they merge into larger rigid clusters. Simultaneously, the power-law exponent shifts closer to the thermodynamic limit (from $-1.92$ to $-1.96$), and the hump itself becomes sharper, higher, and shifts to larger sizes. The strict mutual rigidity condition for merging in rigidity percolation may allow intermediate clusters to persist more readily than in Bernoulli percolation, resulting in their enhanced abundance and altering the power-law decay.

As discussed in Subsection~\ref{subsec:Cascade Dynamics}, merger events in RP can span spatial extents ranging from $R_{\text{rp}}$ to $R_{\text{bp}}$, and the clusters observed in the static distribution are precisely the products of these merger events. We propose that the observed anomalous scaling behavior of the static distribution manifests the interplay of merger processes occurring across this broad range of length scales. The standard FSS form for $P_s$ assumes a single dominant length scale; the violation of this condition in rigidity percolation may lead to the observed deviation for clusters smaller than the cutoff scale. We argue that this does not indicate a genuine departure from $\tau = 2.08$, but rather reflects a slow crossover that requires extremely large system sizes to observe the asymptotic behavior.

\subsection{Dynamic Self-Similarity}
\label{subsec:Dynamic Self-Similarity}

The anomalous scaling behavior observed in static distributions reveals inherent limitations in the analysis of individual time snapshots. To circumvent these constraints and obtain deeper insights into critical phenomena, we examine dynamic distributions that capture the temporal evolution of cluster properties throughout the bond-adding process.

\begin{figure}[tp]
    \centering
    \includegraphics[width=\linewidth]{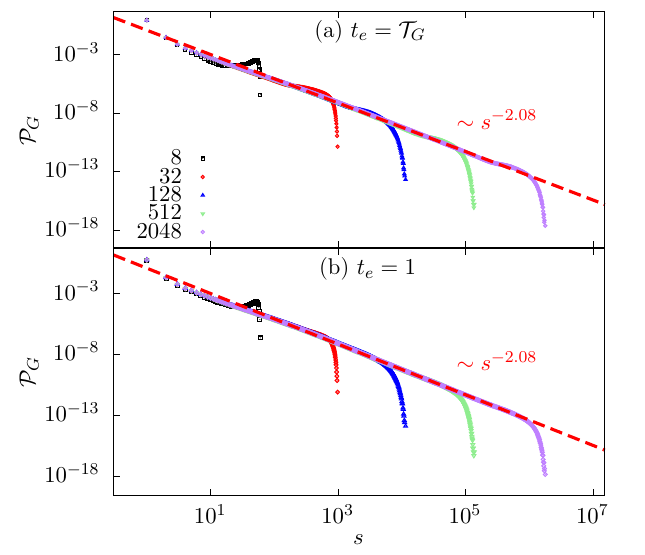}
    \caption{Probability density function of gap sizes ($\mathcal{P}_G$) during the bond-adding process for various system sizes $L$. (a) Distribution measured from empty lattice to pseudo-critical point $\mathcal{T}_G$. (b) Distribution measured from empty lattice to full occupation ($t_e = 1$). Both distributions exhibit clear power-law scaling $\mathcal{P}_G \sim s^{-\tau_g}$ with $\tau_g = \tau = 1 + d/d_f \approx 2.08$. The dashed lines indicate the theoretical slope. The data collapse across system sizes confirms robust dynamic self-similarity in gap statistics.}
    \label{fig:PG}
\end{figure}

\begin{figure}[tp]
    \centering
    \includegraphics[width=\linewidth]{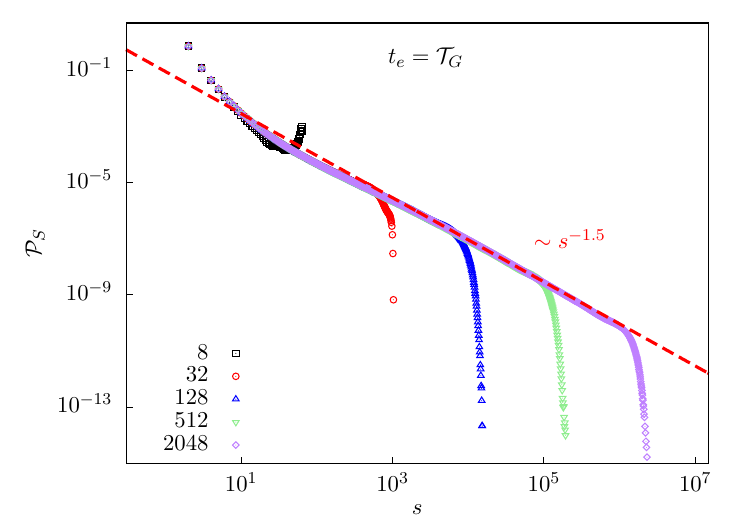}
    \caption{Probability density function of merged cluster sizes ($\mathcal{P}_S$) measured from empty lattice to pseudo-critical point $\mathcal{T}_G$ for various system sizes $L$. The distribution exhibits clear power-law scaling $\mathcal{P}_S \sim s^{-\tau_s}$ with $\tau_s \approx 1.5$. From $1/\nu = 0.850(3)$ and $d_f = 1.850(2)$, one obtains $\sigma = 1/(d_f \nu) \approx 0.459$, yielding $\tau_s = \tau - 1 + \sigma \approx 1.540$. The dashed line indicates the theoretical slope. The data collapse across system sizes confirms robust dynamic self-similarity in merged cluster statistics.}
    \label{fig:PS_tG}
\end{figure}

Figure~\ref{fig:PG} shows the gap size distribution $\mathcal{P}_G$, where the gap measures the size increase of the largest cluster involved in each merger event. The distributions measured from empty lattice configurations to two distinct endpoints—the pseudo-critical point $\mathcal{T}_G$ and complete occupation ($t_e = 1$)—both exhibit clear power-law behavior spanning multiple decades:
\begin{equation}
\mathcal{P}_G(s) \sim s^{-\tau_g},
\end{equation}
establishing dynamic self-similarity for gap statistics. Remarkably, we observe $\tau_g \approx \tau = 1 + d/d_f \approx 1 + 2/1.850 = 2.08$, where $d_f \approx 1.850$ is the fractal dimension from Section~\ref{subsec:df_difficulty}. This correspondence is unexpected within conventional FSS frameworks. We interpret this as evidence that $\mathcal{P}_G$ encodes essential critical behavior when the bond-adding interval encompasses critical or pseudo-critical regions. The gap size $G$ serves as a proxy for the correlation length, governed by a single dominant length scale, thereby yielding clean power-law scaling with exponents consistent with theoretical predictions.

We also examine the merged cluster size distribution $\mathcal{P}_S$, which characterizes the sizes of clusters formed immediately after merger events. As shown in Fig.~\ref{fig:PS_tG} for measurements up to $\mathcal{T}_G$, the distribution follows a power law:
\begin{equation}
\mathcal{P}_S(s) \sim s^{-\tau_s},
\end{equation}
with $\tau_s \approx 1.5$. This exponent follows the relation $\tau_s = \tau - 1 + \sigma$, where $\sigma = 1/(d_f \nu)$ governs the divergence of the cutoff cluster size, $s_{\max} \sim |t - t_c|^{-1/\sigma}$. Employing our values $1/\nu = 0.850(3)$ and $d_f = 1.850(2)$ yields $\tau_s \approx 1.540$, in excellent agreement with observations. This power-law scaling with $\tau_s \approx 1.5$ holds whenever the bond-adding interval encompasses a sufficient portion of the critical region. The detailed derivations of both $\tau_g$ and $\tau_s$ are given in Ref.~\cite{lu2024self}.

The robustness of this dynamic self-similarity emerges through data collapse across diverse system sizes and measurement intervals. Provided the bond-adding range encompasses a sufficient portion of the critical region, power-law scaling invariably emerges, establishing temporal self-similarity as an intrinsic characteristic of rigidity percolation dynamics rather than a finite-size artifact.

\begin{figure}[tp]
    \centering
    \includegraphics[width=\linewidth]{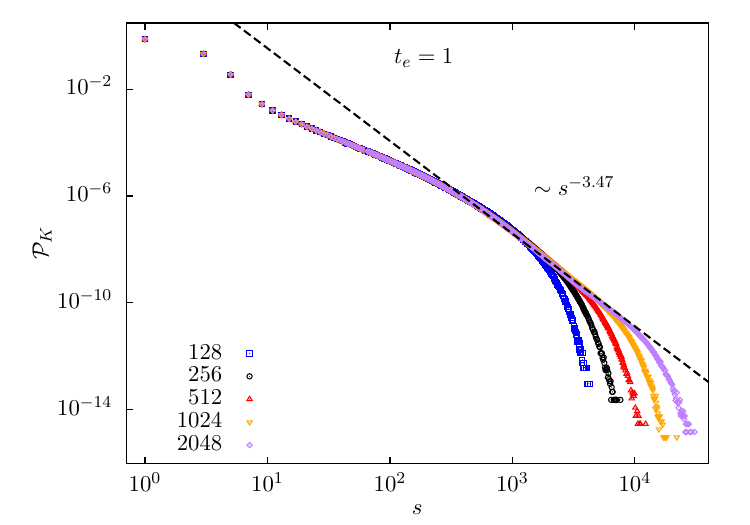}
    \caption{Distribution $\mathcal{P}_{K}$ of the number of merged clusters $\mathcal{K}$ for various system sizes $L$. The distribution is aggregated over the entire bond-adding process. Similar to the cluster growth distribution, the data collapse well and exhibit a clear power-law decay. The dashed line represents a power-law with an exponent of $-3.47$, which describes the data well.}
    \label{fig:pk_dist}
\end{figure}
  
Further intriguing evidence for temporal self-similarity is found in the dynamic distribution of the number of merged clusters, $\mathcal{K}$. As shown in Fig.~\ref{fig:pk_dist}, when aggregated over the entire bond-adding process, the distributions $\mathcal{P}_{K}$ for various system sizes also collapse onto a curve. The data exhibit a robust power-law decay, $\mathcal{P}_{K}(s) \sim s^{-3.47}$, over several decades. Given that the correlation length divergence is driven by the cascade-induced large-scale mergers, the divergence of $K$ reflects that of the correlation length. Numerically, we find that the maximum number of merged clusters scales with an exponent $d_K \approx 1/\nu$, which yields the Fisher exponent $\tau_K = 1 + d/d_K = 1 + d\nu \approx 3.35$, in good agreement with the observed exponent. This indicates that temporal self-similarity is a general feature of the critical dynamics in RP, not restricted to observables related to cluster size or growth. To our knowledge, this is the first report of self-similarity in the distribution of the number of merged clusters for RP.

\subsection{Challenge in determining the fractal dimension}
\label{subsec:df_difficulty}

Rigidity percolation is notorious for its strong finite-size corrections, which render high-precision numerical determination of critical exponents exceedingly challenging. Even with data from large system sizes, up to $L=3200$ in Ref.~\cite{moukarzel1999comparison}, estimates of the fractal dimension $\df$ remain sensitive to the choice of fitting methods. We investigate this issue within our event-based ensembles, as this approach has been shown to effectively suppress finite-size effects in Refs.~\cite{li2023explosive,li2024explosive,li2024crossover}, in the hope of observing cleaner critical behavior. Our analysis yields an improved estimate for $\df=1.850(2)$ with robust stability: the results remain consistent across different fitting methods.

\begin{figure}[tp]
    \centering
    \includegraphics[width=\linewidth]{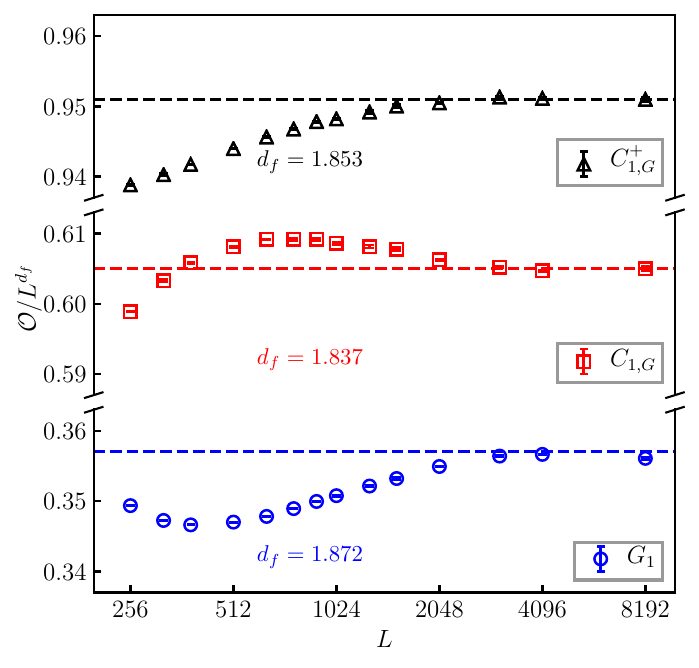}
    \caption{Effective fractal dimension for three observables related to the largest cluster. Here, $C_{1,G}^+$ is the size of the largest cluster immediately after its largest-gap merger, $C_{1,G}$ is its size just before this event, and $G_1$ is the size of the gap. The plot shows the scaled quantities $\mathcal{O}/L^{\df}$ versus system size $L$ for $C_{1,G}^+$ (black triangles), $C_{1,G}$ (red squares), and $G_1$ (blue circles). For each observable, the value of $\df$ is chosen to make the scaled data asymptotically flat at large $L$, giving $\df = 1.853$ for $C_{1,G}^+$, $\df = 1.837$ for $C_{1,G}$, and $\df = 1.872$ for $G_1$, as indicated by the dashed lines. The striking observation is that the three observables require noticeably different values of $\df$ to achieve flat scaling. The data also reveal that the sign of the correction coefficient dominant at small system sizes differs across observables: $G_1$ decreases with increasing $L$ (positive coefficient), while both $C_{1,G}^+$ and $C_{1,G}$ increase with increasing $L$ (negative coefficient). The fluctuations of $C_{1,G}$ and $G_1$ about their respective dashed lines indicate the presence of multiple correction terms with opposite-sign coefficients.}
    \label{fig:df_corrections}
\end{figure}

To probe the system's critical behavior, our analysis focuses on observables defined by the merger across the largest gap of the largest cluster. This event marks a pseudo-critical point $\mathcal{T}_G$ for each dynamic process. We examine the scaling of three quantities: the size of the largest cluster just before the event ($C_{1,G}$), the size of its largest gap ($G_1$), and the size of the new largest cluster immediately after the merger ($C_{1,G}^+$).

\renewcommand{\tablename}{TABLE}
\begin{table*}[tbp]
    \centering
    \caption{Fitting results for the fractal dimension $\df$ from quantities $C_{1,G}^+$, $C_{1,G}$, and $G_1$. Here, $C_{1,G}^+$ is the size of the largest cluster immediately after its largest-gap merger, $C_{1,G}$ is its size just before this event, and $G_1$ is the size of the gap. The results are based on the scaling form in Eq.~\eqref{eq:O_fit}. For each fit, $L_{\mathrm{min}}$ indicates the minimum system size used. Values without error bars are fixed during the fitting procedure. A dash ($-$) indicates that the corresponding term was excluded from the fit. Based on the fitting quality in this table, the finite-size correction analysis in Fig.~\ref{fig:df_corrections}, and the convergence behavior of cluster size distributions across different system sizes in Fig.~\ref{fig:C1_PDFs}, we conclude that the most reliable estimate of the fractal dimension comes from $C_{1,G}^+$, yielding $\df = 1.850(2)$.}
    \label{tab:df_fits}
    \begin{threeparttable}
        \begin{tabular}{c l l l l l l l l}
            \hline\hline
                 & \multicolumn{1}{c}{$L_{\mathrm{min}}$} & \multicolumn{1}{c}{$\df$} & \multicolumn{1}{c}{$a_0$} & \multicolumn{1}{c}{$y_1$} & \multicolumn{1}{c}{$a_1$} & \multicolumn{1}{c}{$y_2$} & \multicolumn{1}{c}{$a_2$} & \multicolumn{1}{c}{$\chi^2/\mathrm{DF}$} \\
            \hline
            \multirow{9}{*}{$C_{1,G}^+$} 
            & 512 & 1.8519(2) & 0.958(1) & $-1$ & $-5.1(2)$ & \multicolumn{1}{c}{$-$} & \multicolumn{1}{c}{$-$} & 17.55/8 \\
            & 768 & 1.8512(3) & 0.963(3) & $-1$ & $-6.3(5)$ & \multicolumn{1}{c}{$-$} & \multicolumn{1}{c}{$-$} & 8.70/6 \\
            & 1024 & 1.8507(4) & 0.967(3) & $-1$ & $-7.4(8)$ & \multicolumn{1}{c}{$-$} & \multicolumn{1}{c}{$-$} & 4.86/4 \\
            & 256 & 1.8445(8) & 1.034(9) & $-0.5$ & $-1.2(1)$ & $-1$ & 5.4(8) & 15.17/9 \\
            & 384 & 1.8465(9) & 1.013(9) & $-0.5$ & $-0.9(1)$ & $-1$ & 3(1) & 6.09/7 \\
            & 512 & 1.847(1) & 1.00(1) & $-0.5$ & $-0.7(2)$ & $-1$ & 1(2) & 5.37/6 \\
            & 512 & 1.8505(4) & 0.969(4) & $-1$ & $-9(1)$ & $-2$ & 1055(306) & 5.77/6 \\
            & 640 & 1.8501(6) & 0.973(5) & $-1$ & $-10(2)$ & $-2$ & 1607(679) & 4.94/5 \\
            & 768 & 1.8500(9) & 0.974(8) & $-1$ & $-11(3)$ & $-2$ & 1929(1340) & 4.83/4 \\
            \hline
            \multirow{7}{*}{$C_{1,G}$} 
            & 512 & 1.8298(3) & 0.643(2) & $-1$ & $-3.2(2)$ & \multicolumn{1}{c}{$-$} & \multicolumn{1}{c}{$-$} & 207.64/8 \\
            & 1024 & 1.8370(7) & 0.603(4) & $-1$ & 5.5(9) & \multicolumn{1}{c}{$-$} & \multicolumn{1}{c}{$-$} & 21.82/4 \\
            & 512 & 1.865(3) & 0.45(1) & $-0.5$ & 2.6(2) & $-1$ & $-25(1)$ & 6.12/6 \\
            & 640 & 1.862(4) & 0.46(2) & $-0.5$ & 2.4(2) & $-1$ & $-23(2)$ & 5.07/5 \\
            & 768 & 1.863(6) & 0.45(3) & $-0.5$ & 2.5(4) & $-1$ & $-24(4)$ & 4.88/4 \\
            & 768 & 1.842(2) & 0.57(9) & $-1$ & 20(4) & $-2$ & $-7173(1534)$ & 6.87/4 \\
            & 1024 & 1.845(2) & 0.56(1) & $-1$ & 28(4) & $-2$ & $-11506(2516)$ & 3.35/3 \\
            \hline
            \multirow{7}{*}{$G_1$} 
            & 512 & 1.8724(5) & 0.361(2) & $-1$ & $-0.34(1)$ & \multicolumn{1}{c}{$-$} & \multicolumn{1}{c}{$-$} & 626.08/8 \\
            & 1024 & 1.853(1) & 0.435(4) & $-1$ & $-1.13(5)$ & \multicolumn{1}{c}{$-$} & \multicolumn{1}{c}{$-$} & 26.91/4 \\
            & 512 & 1.830(2) & 0.56(1) & $-0.5$ & $-3.9(2)$ & $-1$ & 33(2) & 9.21/6 \\
            & 640 & 1.831(3) & 0.55(2) & $-0.5$ & $-3.8(4)$ & $-1$ & 31(4) & 8.84/5 \\
            & 768 & 1.829(5) & 0.56(3) & $-0.5$ & $-3.9(6)$ & $-1$ & 33(7) & 8.47/4 \\
            & 768 & 1.858(2) & 0.409(6) & $-1$ & $-31(3)$ & $-2$ & 8800(1127) & 7.61/4 \\
            & 1024 & 1.855(2) & 0.419(7) & $-1$ & $-37(4)$ & $-2$ & 11991(2063) & 3.72/3 \\
            \hline\hline
        \end{tabular}
    \end{threeparttable}
\end{table*}

To gain initial insight into the fractal dimension and the nature of finite-size corrections, we perform a simple single-parameter analysis. As shown in Fig.~\ref{fig:df_corrections}, we determine for each observable the value of $\df$ that makes the scaled quantity $\mathcal{O}/L^{\df}$ approximately flat at large system sizes. This analysis yields $\df = 1.853$ for $C_{1,G}^+$, $\df = 1.837$ for $C_{1,G}$, and $\df = 1.872$ for $G_1$. Only data with $L \ge 256$ are shown, smaller system sizes are omitted to better visualize the correction behavior at larger system sizes, although they would reveal additional correction terms. A striking observation is that the three observables yield noticeably different values of $\df$. The figure also reveals that the sign of the correction coefficient dominant at small system sizes differs across observables: the scaled data for $G_1$ decrease with increasing $L$ (positive correction coefficient), while those for $C_{1,G}^+$ and $C_{1,G}$ increase with increasing $L$ (negative correction coefficient). Furthermore, the non-monotonic trends—$C_{1,G}$ and $G_1$ first decrease then increase, or vice versa—indicate the presence of multiple correction terms with opposite-sign coefficients.

We now turn to more systematic fits using the form in Eq.~\eqref{eq:O_fit}, with the leading exponent $y_{\mathcal{O}}$ identified as the fractal dimension $\df$, and the results are summarized in Table~\ref{tab:df_fits}. For $C_{1,G}^+$, fits with a single correction term (exponent $y_1 = -1$) yield values of $\df$ that are relatively stable across a wide range of $L_{\mathrm{min}}$, with $\chi^2/\mathrm{DF}$ values approximately equal to one. When we include two correction terms, using either the combination $y_1 = -0.5$ and $y_2 = -1$ or the combination $y_1 = -1$ and $y_2 = -2$, both produce stable fits for $\df$, $a_0$, and $a_1$. Importantly, all three fitting schemes yield consistent results within error bars: the single correction term gives $\df = 1.850(1)$, while the two-correction-term schemes yield $\df = 1.847(1)$ and $\df = 1.850(1)$, respectively. Based on these consistent fits, we obtain $\df = 1.850(2)$ for $C_{1,G}^+$.

For $C_{1,G}$, a single correction term is clearly insufficient. As $L_{\mathrm{min}}$ increases, the fitted value of $\df$ drifts upward, and the sign of the correction coefficient changes from negative to positive. This behavior is consistent with the presence of at least two correction terms with opposite signs, as inferred from Fig.~\ref{fig:df_corrections}. When two correction terms are included, using either the combination $y_1 = -0.5$ and $y_2 = -1$ or the combination $y_1 = -1$ and $y_2 = -2$, both yield stable fits for $\df$, $a_0$, and $a_1$. However, the two correction schemes produce noticeably different estimates: the former gives $\df = 1.865(3)$, while the latter yields $\df = 1.843(2)$.

Similarly, for $G_1$, fits with only one correction term are unstable, with $\df$ decreasing as $L_{\mathrm{min}}$ increases. When two correction terms are included, using either the combination $y_1 = -0.5$ and $y_2 = -1$ or the combination $y_1 = -1$ and $y_2 = -2$, both produce stable fits for $\df$, $a_0$, and $a_1$. Again, the two correction schemes yield markedly different estimates: the former gives $\df = 1.831(3)$, while the latter yields $\df = 1.856(4)$.

These findings highlight the complexity of finite-size corrections for observables related to the largest cluster in rigidity percolation. For $C_{1,G}$ and $G_1$, multiple correction schemes produce stable fits with reasonable statistical quality, yet yield noticeably different estimates of $\df$. Moreover, the correction terms contribute a non-negligible fraction of the total signal, as evidenced by the magnitude of the fitted coefficients $a_1$ and $a_2$ relative to $a_0$. Given these differences across the three observables and their respective fitting schemes, it is important to assess which estimate provides the most reliable determination of $\df$.

As shown in Table~\ref{tab:df_fits}, $C_{1,G}^+$ requires only a single correction term to yield stable fits with good $\chi^2/\mathrm{DF}$ values. In contrast, both $C_{1,G}$ and $G_1$ require two correction terms to achieve comparable fit quality. An intriguing observation emerges when examining the two correction schemes that produce reliable fits for $C_{1,G}$ and $G_1$: both correction coefficients for $C_{1,G}$ and $G_1$ have opposite signs, and their absolute values are approximately equal for the subleading terms. Since $C_{1,G}^+ = C_{1,G} + G_1$, the cancellation of both correction terms, particularly the subleading one, explains why $C_{1,G}^+$ exhibits small finite-size corrections and simpler scaling behavior. As noted earlier, the three fitting schemes for $C_{1,G}^+$ yield consistent results within error bars. In contrast, $C_{1,G}$ and $G_1$ exhibit complex correction terms, and their fitted $\df$ values vary noticeably across different correction schemes. We therefore adopt the effective exponent of $C_{1,G}^+$ as our estimate of the fractal dimension, $\df = 1.850(2)$, where the uncertainty reflects the variation across different correction schemes.

Based on the detailed analysis of configurations before and after $\mathcal{T}_G$, we can now provide insight into the origin of the strong finite-size corrections in rigidity percolation. In Bernoulli percolation, the merger of two clusters is essentially additive; even the largest-gap merger, which corresponds to the coalescence of the largest cluster $C_1$ with $C_{2,m}$ (the maximum of $C_2(T)$ over the entire bond-adding process)~\cite{nagler2011impact}, does not change the scaling of $C_1$, because the fractal dimension of $C_{2,m}$ is less than or equal to that of $C_1$. In contrast, rigidity percolation features multi-cluster mergers near the critical point, with some involving numbers of clusters that scale with system size with exponent $1/\nu=0.850(3)$. Since the gap equals the sum of all participating cluster sizes minus the largest one and shared pivots are counted only once, its scaling involves multiple correction terms. This is consistent with Fig.~\ref{fig:df_corrections}, which shows that a large value of $\df \approx 1.872$ is required to make $G_1/L^{\df}$ approximately flat at large $L$, and the non-monotonic trend indicates multiple correction terms.

$C_{1,G}$, formed through multiple large-scale mergers, inherits complex finite-size corrections from the gaps of previous merger events. As shown in Table~\ref{tab:df_fits}, both $C_{1,G}$ and $G_1$ can be fitted using the same correction term combinations, supporting this interpretation. The fact that $C_{1,G}^+$ exhibits simpler scaling is a fortunate consequence of the partial cancellation of correction terms from its constituent parts. This finding also explains the difficulties encountered in traditional fixed-probability ensembles. Such ensembles average over configurations that, due to fluctuations of the pseudo-critical point, may lie either before or after $\mathcal{T}_G$. Consequently, the ensemble-averaged largest cluster, $C_{1,t_c}$, is a mixture of pre- and post-$\mathcal{T}_G$ states; this mixing complicates its finite-size corrections. Since the average pseudo-critical point typically lies below $t_c$, post-$\mathcal{T}_G$ configurations are sampled more often, and $C_{1,t_c}$ primarily inherits the scaling behavior of $C_{1,G}^+$, leading to its effective fractal dimension close to that of $C_{1,G}^+$.

\begin{figure}[tp]
    \centering
    \includegraphics[width=\linewidth]{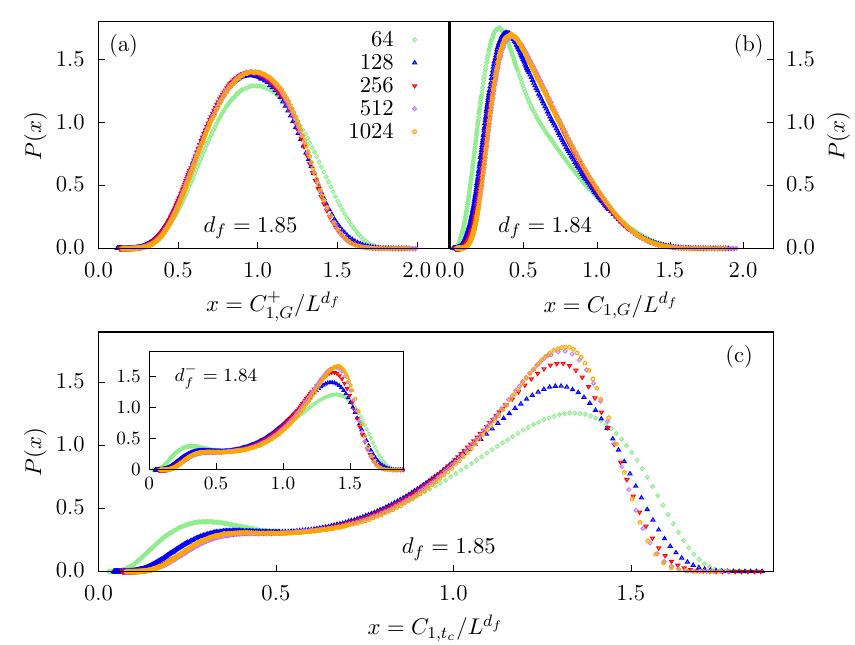}
    \caption{Probability density functions of the largest cluster size in different ensembles. (a) $C_{1,G}^+$ immediately after $C_1$'s largest gap merger event, scaled with $\df = 1.85$. (b) $C_{1,G}$ immediately before the merger, scaled with $\df = 1.84$. (c) $C_{1,t_c}$ in the traditional fixed-probability ensemble at $t_c$, scaled with $\df = 1.85$. The inset shows the same data with $\df = 1.84$. The bimodal structure in panel (c) reflects the mixture of pre- and post-$\mathcal{T}_G$ configurations in the traditional ensemble. Comparing the three panels, the excellent data collapse in panel (a) across system sizes from $L = 256$ to $L = 1024$ indicates weak finite-size corrections, making $C_{1,G}^+$ more suitable than the other two quantities for determining the fractal dimension.}
    \label{fig:C1_PDFs}
\end{figure}

We also investigate the probability density functions (PDFs) of $C_1$ in different ensembles to gain further insight into the largest cluster behavior. Figure~\ref{fig:C1_PDFs} shows the distributions for three cases: (a) $C_{1,G}^+$, (b) $C_{1,G}$, and (c) $C_{1,t_c}$ in the traditional fixed-probability ensemble at $t_c$. At small system sizes, all three distributions exhibit pronounced finite-size corrections. Data collapse is achieved for $L \ge 128$ in case (a), $L \ge 256$ in case (b), while even at $L = 1024$, the right peak in case (c) fails to collapse. The distributions of $C_{1,G}^+$ and $C_{1,G}$ each display a single peak, consistent with standard FSS expectations. However, achieving data collapse requires different values of $\df$ for each quantity, consistent with our results from Table~\ref{tab:df_fits} and Fig.~\ref{fig:df_corrections}. In contrast, the PDF of $C_{1,t_c}$ exhibits a bimodal structure with two distinct peaks. Using the same fractal dimension as for $C_{1,G}^+$ ($\df = 1.85$) produces approximate data collapse for the right peak, although the collapse remains imperfect even for the largest system sizes ($L = 512$ and $1024$), and attempts with other values of $\df$ do not yield better results. Moreover, this peak is located at larger scaled values than those observed for $C_{1,G}^+$, indicating contributions from merger events occurring at bond densities $t > \mathcal{T}_G$, where subsequent large-scale mergers produce discernible changes in the cluster size. The inset of panel (c) shows the same data rescaled with $\df = 1.84$. Here, the left peak collapses well across different system sizes, and the peak position (in terms of the scaled variable $x$) closely matches that of $C_{1,G}$ in panel (b). This demonstrates that the left peak in the $C_{1,t_c}$ distribution originates from configurations sampled before $C_1$'s largest gap merger event. These observations directly confirm that $C_{1,t_c}$ represents a mixture of pre- and post-$\mathcal{T}_G$ states. Comparing the quality of data collapse, $C_{1,G}^+$ shows good collapse already at $L = 128$, while $C_{1,G}$ exhibits notable deviation at this size, and $C_{1,t_c}$ shows bimodal structure that fails to collapse under any single value of $\df$, demonstrating that $C_{1,G}^+$ has the weakest finite-size corrections and is the most suitable quantity for determining the fractal dimension.

\begin{figure}[tp]
    \centering
    \includegraphics[width=\linewidth]{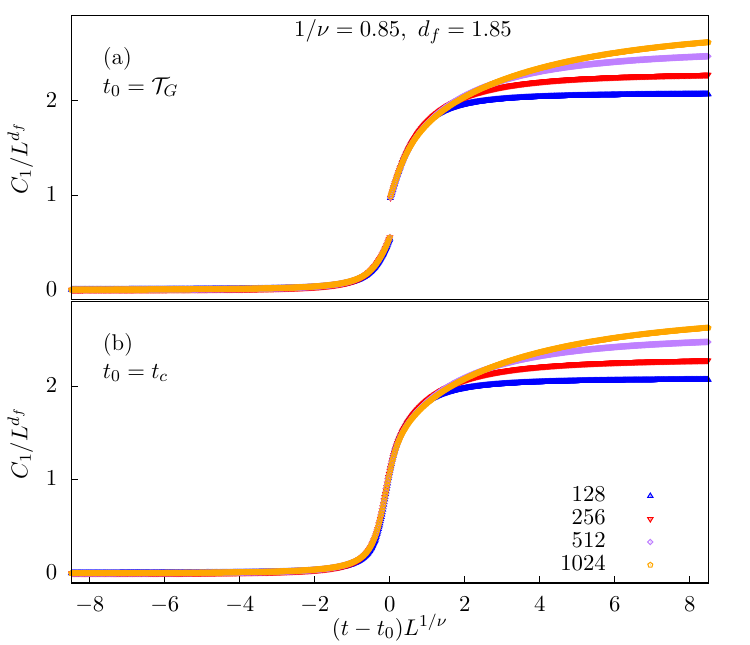}
    \caption{Critical window scaling of the largest cluster size $C_1/L^{\df}$ versus the rescaled time $(t-t_0)L^{1/\nu}$ with $1/\nu = 0.850$ and $\df = 1.850$. (a) Critical window centered on the gap pseudo-critical point $\mathcal{T}_G$. (b) Critical window centered on the critical point $t_c = 0.660\,277\,8$. The discontinuity in panel (a) reflects the merger event at $\mathcal{T}_G$. The critical window is notably narrow, with data collapse limited to a small range of $|(t-t_0)L^{1/\nu}| \lesssim 3$ before curves from different system sizes begin to diverge. This narrow critical window reflects the large-scale cluster merger events near criticality.}
    \label{fig:critical_window}
\end{figure}

Given that different values of $\df$ are required for data collapse of $C_{1,G}^+$ and $C_{1,G}$, we examine the critical window behavior—the range over which FSS theory with a single set of exponents can describe the system. We investigate the scaling collapse of $C_1$ within the critical region. Figure~\ref{fig:critical_window} shows the scaled largest cluster size as a function of the rescaled time variable $(t-t_0)L^{1/\nu}$, where $t_0$ represents either $\mathcal{T}_G$ or $t_c$, using the fitted values $1/\nu = 0.850$ and $\df = 1.850$. In the critical window centered on $\mathcal{T}_G$ [panel (a)], a distinct discontinuity is observed across the merger event, manifesting as a jump in the scaling function. In Bernoulli percolation, such a gap also occurs at $C_1$'s largest gap merger, but since the fractal dimension remains unchanged before and after the merger, a single set of scaling exponents achieves data collapse across the entire critical window. Here, both panels demonstrate that the chosen exponents produce reasonable collapse of data from different system sizes. However, the critical window in rigidity percolation is notably narrow on both sides of the reference points. The range over which curves from different system sizes collapse consistently is limited, extending only to $|(t-t_0)L^{1/\nu}| \lesssim 3$ before curves begin to diverge. This narrow critical window arises from the large-scale cluster merger events near criticality, which drive rapid changes in the effective critical exponents and thereby restrict the region where the set of critical exponents remains valid. 

\section{Conclusion}
\label{Sec:Conclusion}

We have developed a dynamic PG algorithm that enables a comprehensive analysis of rigidity percolation through the sequential addition of bonds to an empty lattice. This computational framework provides detailed access to the evolution of rigid clusters, gap sizes, and merger dynamics at each time step. The algorithm maintains comparable computational efficiency to traditional static PG methods in terms of system size scaling, enabling large-scale simulations up to $L = 8192$. Most notably, our numerical investigations uncover a previously overlooked temporal self-similarity in rigidity percolation, manifested through clean power-law scaling in the distributions of both gap sizes ($\mathcal{P}_G$) and merged cluster sizes ($\mathcal{P}_S$). This temporal scaling exhibits reduced finite-size corrections compared to the anomalous scaling observed in static cluster size distributions, offering a complementary perspective on the critical behavior. Complementing this, we demonstrate that the maximum number of clusters merged by a single bond addition scales with system size, establishing the divergent nature of these large-scale cascade events, whose statistics also exhibit temporal self-similarity.

Leveraging the dynamic algorithm, we achieve high-precision estimates of the critical point and critical exponents through event-based ensembles that allow critical behavior measurements without prior knowledge of the critical point. We employ a pseudo-critical point $\mathcal{T}_\chi$ defined by the maximum susceptibility, which exhibits simple FSS with relatively weak corrections, enabling us to improve the precision of the critical point estimate by three orders of magnitude to $t_c = 0.660\,277\,8(10)$ and extract the inverse correlation-length exponent $1/\nu = 0.850(3)$. Furthermore, a detailed FSS analysis within an event-based ensemble centered on the largest-gap merger of the largest cluster ($\mathcal{T}_G$) yields the fractal dimension $d_f = 1.850(2)$ from the scaling behavior of the largest cluster after the merger ($C_{1,G}^+$), where the finite-size corrections for $C_1$ are reduced compared to traditional ensembles. This analysis reveals that the strong finite-size corrections in rigidity percolation originate from the merger of an extensive number of clusters.

The number of clusters merged by a single bond can vary over a broad range, with the spatial extent of merger events fluctuating between scales from $R_{\text{rp}}$ to $R_{\text{bp}}$. Near criticality, the maximum number of merged clusters scales with system size, enabling large-scale merger events that incorporate numerous clusters with fractal dimensions spanning a wide range. This incorporation of clusters with diverse fractal dimensions contributes to substantial correction terms in the finite-size scaling and may potentially elevate the effective fractal dimension above that of the largest pre-merger constituent, resulting in the exceptionally large finite-size corrections observed in RP. These observations hint at the possible existence of multiple divergent correlation length scales in rigidity percolation.

In contrast to the anomalous scaling observed in static cluster distributions, the clean power-law behavior of both dynamic gap distributions $\mathcal{P}_G$ and merged cluster size distributions $\mathcal{P}_S$ provides insights into the temporal self-similarity of rigidity percolation. This temporal scaling exhibits reduced finite-size corrections compared to spatial distributions. Utilizing the scaling forms $\mathcal{P}_G(s) \sim s^{-\tau_g}$ and $\mathcal{P}_S(s) \sim s^{-\tau_s}$, where $\tau_g = \tau = 1 + d/d_f$ and $\tau_s = \tau - 1 + \sigma = 1 + d/d_f - 1 + 1/(d_f \nu)$, the fractal dimension $d_f$ and correlation length exponent $\nu$ can be determined by measuring the fraction of gaps exceeding size thresholds and analyzing their scaling behavior through systematic FSS analysis.

The event-based ensemble approach has proven effective in bond rigidity percolation, enabling high-precision determination of the critical point, correlation length exponent $\nu$, and fractal dimension $d_f$. This methodology can be readily extended to site rigidity percolation. Beyond critical exponent determination, the detailed measurements of merger dynamics afforded by the dynamic algorithm offer promising directions for further research to illuminate the distinctions between RP and BP. Moreover, while RP exhibits a continuous phase transition in two dimensions, it undergoes a first-order transition on complete graphs. The dynamic algorithm framework can help investigate the crossover from continuous to first-order transitions as the system dimensionality increases from two dimensions to high dimensions.

\section{Acknowledgements}

We thank Xiaojun Tan for his initial involvement on the code for static pebble-game algorithm. This work has been supported by the National Natural Science Foundation of China (under Grant No.12275263), the Innovation Program for Quantum Science and Technology (under grant No. 2021ZD0301900).

 \bibliographystyle{apsrev4-2}
 \bibliography{ref1.bib}

 \appendix

\section{Pseudo-Critical Point Data}
\label{app:pseudo_critical_data}

Table~\ref{tab:tK_tchi_data} presents the values of the pseudo-critical points $t_K$ and $t_{\chi}$ along with their statistical uncertainties for selected system sizes. These data are used to analyze the finite-size scaling behavior discussed in Sec.~\ref{Sec:Results}.

\renewcommand{\tablename}{TABLE}
\begin{table}[tbp]
    \centering
    \caption{Values of pseudo-critical points $t_G$, $t_{\chi}$, and $t_K$ with their statistical uncertainties for different system sizes $L$. The uncertainties are indicated in parentheses and correspond to the Monte Carlo error in the last digit(s).}
    \label{tab:tK_tchi_data}
    \begin{threeparttable}
        \begin{tabular}{l l l l}
            \hline\hline
            \multicolumn{1}{c}{$L$} & \multicolumn{1}{c}{$t_G$} & \multicolumn{1}{c}{$t_{\chi}$} & \multicolumn{1}{c}{$t_K$} \\
            \hline
            8 & 0.629017(5) & 0.61037(2) & 0.630444(5) \\
            16 & 0.648315(3) & 0.63619(1) & 0.650203(3) \\
            32 & 0.653940(2) & 0.646988(6) & 0.656323(2) \\
            64 & 0.656100(2) & 0.652749(3) & 0.658619(2) \\
            128 & 0.6575647(9) & 0.656043(2) & 0.659570(1) \\
            256 & 0.6586926(5) & 0.657914(1) & 0.6599724(9) \\
            512 & 0.6593913(3) & 0.6589638(5) & 0.6601359(5) \\
            1024 & 0.6597863(4) & 0.6595480(3) & 0.6601977(6) \\
            2048 & 0.6600060(1) & 0.6598729(2) & 0.660257(2) \\
            4096 & 0.6601272(1) & 0.6600535(2) & 0.6602435(1) \\
            8192 & 0.6601937(1) & 0.6601529(1) & 0.6602557(2) \\
            \hline\hline
        \end{tabular}
    \end{threeparttable}
\end{table}

\section{The Dynamic Pebble Game Algorithm}
\label{app:dynamic_pebble_game}

The rigidity percolation simulations presented in this work employ an event-based ensemble approach, where bonds are added sequentially to an initially empty triangular lattice. This methodology requires an efficient algorithm capable of tracking the evolution of rigid clusters after each bond addition. While the standard PG algorithm~\cite{jacobs1996generic} provides a robust framework for analyzing the rigidity of static networks, its direct application to dynamic simulations poses computational challenges that necessitate algorithmic innovations.

\subsection{Computational Challenges and Motivation}

The naive approach of applying the standard PG algorithm after each bond addition encounters severe computational bottlenecks. While traditional fixed-probability ensembles perform cluster identification only once after all bonds are placed, dynamic simulations require cluster re-identification after every bond addition. The core computational challenge arises from cluster merger events: when multiple rigid clusters merge, the algorithm must traverse all vertices belonging to the merging clusters and their neighbors to ensure that no clusters that should be merged are overlooked in determining the new rigid cluster.

The computational burden becomes particularly severe near and above the critical point, where large rigid clusters emerge and dominate the system. In the supercritical phase, the largest cluster reaches a size of $O(L^{2})$, and merger events involving these giant clusters occur with a frequency of $O(L^2)$ as bonds are added from the critical point to full occupation. Each such merger requires traversing the entire giant cluster and its neighbors, leading to a computational complexity of $O(L^{2} \times L^2) = O(L^{4})$ for the supercritical phase alone. This is substantially higher than the $O(L^{2.3})$ complexity of traditional fixed-probability ensemble simulations and renders large-scale dynamic simulations impractical without algorithmic innovations.

To overcome these limitations, we developed a dynamic PG algorithm that maintains persistent data on cluster membership. Specifically, the algorithm stores two mappings at each time step: one associating each vertex with the set of clusters to which it belongs, and another associating each cluster with the set of vertices it contains. This approach, in contrast to naive implementations that discard such information, yields two computational advantages: first, it enables rapid determination of bond independence without requiring pebble searches; second, it substantially accelerates the cluster merging process. The speedup is achieved by excluding the largest cluster involved in a merger from the traversal process, which significantly reduces computational cost while ensuring the correct identification of the final merged cluster.

\subsection{Theoretical Foundation}

Our algorithm achieves efficiency by leveraging stored cluster membership information. The key insight for rapid bond independence determination is straightforward: two vertices are mutually rigid if and only if they belong to the same rigid cluster. When a new bond connecting vertices $u$ and $v$ is added, if they already belong to the same rigid cluster, they are mutually rigid, making the bond redundant. Conversely, if they belong to different clusters (or no cluster), they are not mutually rigid, and the bond is independent.

To establish the theoretical foundation for our cluster merging optimizations, we first clarify some terminology. In rigidity percolation, a rigid cluster is the maximal connected component of mutually rigid vertices and their connecting bonds. A graph $G$ is a set of vertices $V$ and edges $E$. Importantly, multiple rigid graphs can belong to the same rigid cluster, but different rigid clusters are, by definition, not mutually rigid.

Consider two rigid graphs $G_1$ and $G_2$ with $N_1$ and $N_2$ vertices, $E_1$ and $E_2$ independent bonds, respectively, and $n$ shared vertices. This analysis focuses on the case where only these two graphs are involved. For each graph to be rigid, we have the constraints $2N_i - E_i = 3$ (where $i = 1, 2$). When combined, the total number of vertices is $N_1 + N_2 - n$ and the total number of independent bonds is $E_1 + E_2$. The degrees of freedom (DOF) of the combined system is:
\begin{equation}
    \text{DOF} = 2(N_1 + N_2 - n) - (E_1 + E_2) = 6 - 2n.
\end{equation}

To examine the overall rigidity of the combined graph, we require DOF $\leq 3$. This yields $n \geq 1.5$, which leads to $n \geq 2$ for integer $n$. This analysis establishes:

\begin{description}
    \item[Lemma 1:] Two rigid graphs sharing exactly one vertex cannot form a rigid total graph.
    \item[Lemma 2:] Two rigid graphs sharing two or more vertices can form a rigid total graph.
\end{description}

From these lemmas, we derive a third result for cluster merging scenarios:

\begin{description}
    \item[Lemma 3:] In any cluster merger involving multiple clusters, each participating cluster must share vertices with at least two other clusters in the merger set.
\end{description}

\textit{Proof:} If a cluster shares one vertex with only one other cluster, it cannot merge with additional clusters (Lemma 1). If it shares two or more vertices with any single cluster, they would already constitute the same rigid entity (Lemma 2), contradicting the assumption of multiple distinct clusters. Therefore, each cluster must have single-vertex connections with at least two other clusters.

These lemmas enable the \textbf{let go of the largest} trick: during the traversal-based merger identification, we can safely omit one cluster from explicit verification—that is, we need not traverse all neighbors of its vertices—without missing any mergers. All mergeable clusters will be encountered through the remaining clusters' neighborhoods, as guaranteed by Lemma 3. Therefore, we can confidently exclude the largest cluster in the merger set from neighbor traversal, substantially reducing the computational complexity of cluster merging.

An additional optimization applies when either vertex of a new bond belongs to no existing cluster. In such cases, we need not search for multi-cluster mergers. If both vertices belong to no clusters, the new bond forms an isolated cluster, naturally requiring no merger checks with other clusters. If only one vertex belongs to existing clusters, then the initial rigid graph $G_0$ (formed by the new bond and its endpoints) shares only one vertex with those clusters, and by Lemma 1, it cannot merge with other clusters.

\subsection{Core Data Structures}

The efficiency of our dynamic approach relies on specialized data structures that maintain real-time information about the relationship between vertices and rigid clusters. Unlike traditional PG implementations, which perform cluster identification only after all bonds are placed and thus do not require persistent cluster tracking, our dynamic algorithm maintains three key data structures throughout the bond-adding process. The \textbf{Cluster Table} stores, for each cluster, the first and last sites added to that cluster. The \textbf{Site Table} records, for each vertex, the clusters it belongs to (identified by root bond numbers) and the next site in the cluster's site chain. Additionally, a \textbf{Parent Array} implements a union-find data structure to efficiently manage cluster information during mergers.

\subsubsection{The Cluster Table}

The Cluster Table, denoted as $\mathcal{C}$, has size $(B, 2)$, where $B = 3L^2$ is the total number of bonds. For each bond, it stores the head and tail of the site chain within the cluster, corresponding to the first and last vertices added to that cluster. This enables convenient traversal of cluster members and efficient insertion of new sites at the tail.

The Cluster Table is used after rigidity has been established (via PG algorithms) to traverse the vertices of rigid clusters and search for neighboring clusters that might be mutually rigid. The advantage of this structure is that it allows direct traversal of all vertices within a cluster without requiring neighbor searches, improving computational efficiency during potential cluster merger identification.

\subsubsection{The Site Table}

The Site Table, denoted as $\mathcal{S}$, has size $(L^2, 0:\text{nnb}, 2)$, where $L^2$ corresponds to all vertices and nnb represents the maximum number of clusters to which a vertex belongs (at most 6 in the triangular lattice). The index 0 stores the number of clusters to which the vertex belongs. For each cluster membership, the table stores the root node of that cluster and the index of the next vertex in the site chain (with 0 indicating the chain's tail).

An important advantage of the Site Table is that traversing the clusters to which a vertex belongs does not require traversing all neighbors to determine cluster membership. Instead, we can directly iterate through the clusters to which a vertex belongs, with the number of iterations bounded by 6, which is typically smaller than or equal to the coordination number.

\subsubsection{The Parent Array}

The Parent Array has size $B$ and implements a union-find data structure, following established methods for efficient cluster management in percolation simulations~\cite{newman2000efficient, newman2001fast, tan2020n}. If a bond is unoccupied, its value is 0. If occupied but not a root node, it points to a parent or root node in the cluster tree. If it is a root node, it stores the negative of the cluster size (defined by vertex count), so a cluster with three vertices has value $-3$. This structure is highly efficient, enabling nearly constant-time (amortized) cluster merging operations. Additionally, it provides direct access to cluster sizes, which is used for implementing the \textbf{let go of largest} trick where the largest cluster is identified and skipped during traversal.

\subsection{Dynamic Update Procedure}

The algorithm processes each bond addition through a two-stage procedure designed to optimize computational efficiency while accurately obtaining complete information about all rigid clusters at each time step.

\subsubsection{Stage 1: Efficient Bond Independence Classification}

When a new bond connecting vertices $u$ and $v$ is added to the system, the first step is to determine whether this bond is independent (contributes to system rigidity) or redundant (does not alter the rigid cluster structure). Traditional PG implementations perform this classification by searching for four free pebbles around the new bond, a process that may require extensive graph traversal.

Our dynamic algorithm exploits the maintained cluster membership information to perform this classification efficiently:

\begin{enumerate}
    \item Retrieve cluster memberships: Check $\mathcal{S}(u)[0]$ and $\mathcal{S}(v)[0]$ to determine the number of clusters each vertex belongs to.
    \item Handle trivial cases: If either vertex belongs to no rigid clusters ($\mathcal{S}(u)[0] = 0$ or $\mathcal{S}(v)[0] = 0$), the bond is immediately classified as independent.
    \item Check for shared clusters: Iterate through the cluster root nodes stored in $\mathcal{S}(u)$ and $\mathcal{S}(v)$ to determine if they share any common cluster.
    \item Classify bond:
    \begin{itemize}
        \item If no shared clusters are found, vertices $u$ and $v$ are not mutually rigid. The bond is \textbf{independent}.
        \item If any shared cluster is found, vertices $u$ and $v$ are already connected through a rigid cluster. The bond is \textbf{redundant}.
    \end{itemize}
\end{enumerate}

This classification method requires only set operations on small collections (maximum six elements per vertex) and avoids the graph traversal inherent in pebble searching. The computational complexity is reduced to $O(1)$ for our approach.

\subsubsection{Stage 2: Cluster Merging with the \textbf{let go of the largest} trick}

When an independent bond is identified, it may trigger the merger of previously distinct rigid clusters. The challenge lies in efficiently determining which clusters participate in the merger and updating the data structures accordingly.

Our algorithm employs a cluster-based traversal strategy rather than the vertex-based approach used in standard implementations. The procedure begins by treating the newly added bond and its endpoints as an initial rigid cluster. All existing clusters to which these endpoints belong are candidates for merger and are placed in a processing queue.

The algorithm then iterates through the queue, processing each candidate cluster:

\begin{enumerate}
    \item Dequeue a cluster $C_i$ from the processing queue.
    \item Using the Cluster Table $\mathcal{C}$, traverse all unvisited vertices in $C_i$ by following the site chain from head to tail, marking each vertex as visited.
    \item For each vertex $v$ in the cluster, check the clusters it belongs to using $\mathcal{S}(v)$.
    \item For each unvisited cluster that vertex $v$ belongs to, perform a pebble search to determine if it is rigid with respect to the newly added bond, then mark this cluster as visited.
    \item If rigid connection is confirmed, add the cluster to the merger set and to the processing queue.
    \item Continue until the queue is empty.
\end{enumerate}

Upon completion, all clusters visited during this process are merged into a single rigid cluster, and both data structures $\mathcal{C}$ and $\mathcal{S}$ are updated to reflect the new cluster structure.

A key optimization, the \textbf{let go of the largest} trick, dramatically improves computational efficiency by exploiting the theoretical foundation established by Lemma 3. While all participating clusters must undergo rigidity verification during merger events, the neighbor traversal process can be optimized by excluding one cluster from explicit traversal.

Since merger events involve only odd numbers of clusters (see Appendix~\ref{app:odd_merger_proof}), any merger must include at least three clusters. The algorithm dynamically identifies the largest participating cluster and excludes it from neighbor traversal. As additional clusters are discovered during the merging process, their sizes are compared with the current largest cluster. If a newly identified cluster exceeds the size of the current largest, it becomes the new largest cluster, and the previous largest cluster is then traversed to ensure completeness of the merger identification.

This strategy provides substantial computational savings both near the critical point and in the supercritical region. Near criticality, the largest cluster reaches size $O(L^{d_f})$, while in the supercritical phase, it grows to $O(L^2)$. Since merger events frequently involve these largest clusters, avoiding their traversal eliminates the most computationally expensive component of the update process. The \textbf{let go of the largest} trick reduces the computational complexity for bond-addition simulations from subcritical to full occupation from worse than $O(L^4)$ to $O(L^{2.3})$, enabling simulations at system sizes comparable to those achievable with fixed-probability ensemble methods.

\subsection{Performance and Scalability}

\begin{figure}[tp]
    \centering
    \includegraphics[width=\linewidth]{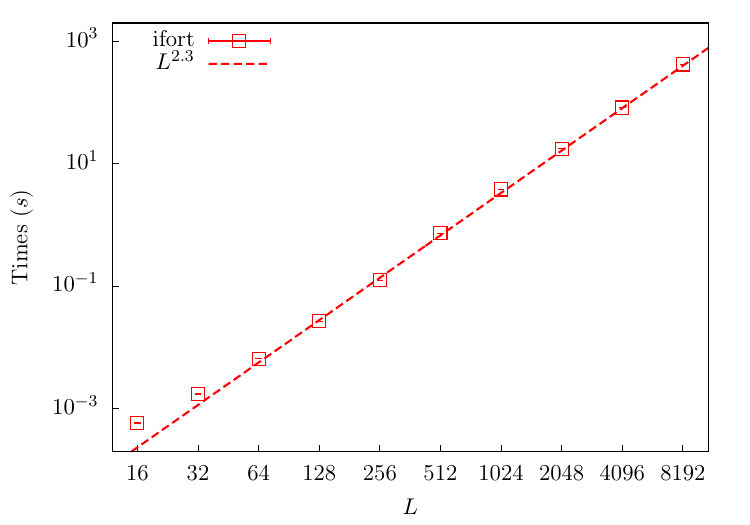}
    \caption{Computational time scaling of the dynamic PG algorithm. The data show the wall-clock time required to simulate a single configuration from empty lattice to full occupation, including measurements of pseudo-critical points and the largest cluster throughout the process. The dashed line represents $L^{2.3}$ scaling, demonstrating that the dynamic algorithm achieves computational efficiency comparable to traditional PG algorithms in fixed-probability ensembles. Simulations were performed using ifort compiler with default optimization on a node with four Intel Xeon Gold 6448H CPUs (128 cores, fully utilized).}
    \label{fig:performance}
\end{figure}

The dynamic PG algorithm presented here successfully addresses the computational bottleneck of repeated rigid cluster identification required after each bond addition. While cluster merger verification in the dynamic algorithm remains more time-consuming than in static methods, the combination of persistent data structures (Cluster Table, Site Table, and Parent Array) and the \textbf{let go of the largest} optimization substantially reduces computational overhead by eliminating the need for PG searches in bond independence classification. Overall, the algorithm achieves computational efficiency comparable to traditional PG algorithms in fixed-probability ensembles. As shown in Fig.~\ref{fig:performance}, benchmarks on a cluster node equipped with four Intel Xeon Gold 6448H CPUs (128 cores total, fully utilized) demonstrate that the algorithm scales as $L^{2.3}$ across system sizes from $L = 16$ to $L = 8192$, with a single configuration (including measurements of pseudo-critical points and the largest cluster $C_1$ throughout the bond-adding process) requiring approximately 17.8 seconds for $L = 2048$ and 421 seconds for $L = 8192$. This scaling behavior ensures that simulations remain computationally feasible even at $L = 8192$, enabling the large-scale investigations necessary for high-precision critical exponent determination while providing direct access to cluster information at each time step and enabling detailed investigation of the cluster merger process.

\section{Merger Cluster Numbers Must Be Odd}
\label{app:odd_merger_proof}

In rigidity percolation, the constraints imposed by mechanical stability restrict the possible configurations of rigid clusters. These constraints affect both the structure of individual clusters and the nature of merger events, creating characteristics that differentiate rigidity percolation from standard Bernoulli percolation.

Consider the merger of $K$ rigid clusters with sizes $N_1, N_2, \ldots, N_K$ (measured in terms of vertices) and independent edge counts $E_1, E_2, \ldots, E_K$. Let $n$ denote the total number of shared vertices among all clusters. For each individual cluster to be rigid, the constraint $2N_i - E_i = 3$ must be satisfied. When these clusters merge, the resulting structure has $N_{\text{total}} = \sum_i N_i - n$ vertices and $E_{\text{total}} = \sum_i E_i$ independent edges.

For the merged structure to be rigid, we require:
\begin{equation}
2N_{\text{total}} - E_{\text{total}} = 3
\end{equation}

Substituting the expressions above:
\begin{equation}
2\left(\sum_i N_i - n\right) - \sum_i E_i = 3
\end{equation}

Using the rigidity constraint for individual clusters, $\sum_i (2N_i - E_i) = 3K$, we obtain:
\begin{equation}
3K - 2n = 3
\end{equation}

This simplifies to $n = \frac{3(K-1)}{2}$. For $n$ to be an integer, $K$ must be odd.

This mathematical constraint helps to explain the pronounced finite-size effects observed in rigidity percolation. Unlike in Bernoulli percolation, where clusters of any bond count can exist, the merger rules in rigidity percolation forbid the formation of certain small cluster sizes. For instance, on the triangular lattice, rigid clusters with exactly 4 or 6 bonds cannot be formed from the merger of an odd number of smaller rigid clusters. While three single-bond clusters can merge into a three-bond cluster, no valid combination of clusters with 1 or 3 bonds can produce a cluster with 4 or 6 bonds. This restriction on the existence of specific small cluster sizes alters the cluster probability distribution, which is a contributing factor to the strong finite-size corrections in the model.

\end{document}